\documentclass[10pt]{article}
\usepackage[utf8]{inputenc}
\usepackage[margin=1in]{geometry}

\usepackage{hyperref}       
\usepackage{url}            
\usepackage{booktabs}       
\usepackage{amsfonts}       
\usepackage{nicefrac}       
\usepackage{microtype}      
\usepackage{graphicx}
\usepackage{subcaption}
\usepackage{booktabs} 
\usepackage{amsmath,amsfonts,amssymb,amsthm}
\usepackage{amsfonts}
\usepackage{dsfont}
\usepackage{xcolor}
\usepackage{mathrsfs}
\usepackage{centernot}
\usepackage{mathtools}
\usepackage{mathrsfs}
\usepackage{wrapfig}
\usepackage{authblk}

\usepackage{soul}
\usepackage{arydshln}
\usepackage[flushleft]{threeparttable}

\title{High-resolution Spatio-temporal Model for\\County-level COVID-19 Activity in the U.S.}
\author[1]{Shixiang Zhu}
\author[1]{Alexander Bukharin}
\author[1]{Liyan Xie}
\author[2]{Mauricio Santillana}
\author[1]{Shihao Yang}
\author[1]{Yao Xie}
\affil[1]{School of Industrial and Systems Engineering, Georgia Institute of Technology}
\affil[2]{Harvard Medical School, Computational Health Informatics Program, Harvard University School of Engineering \& Applied Sciences}

\date{}  

\begin{document}

\maketitle

\begin{abstract}
We present an interpretable high-resolution spatio-temporal model to estimate COVID-19 deaths together with confirmed cases one-week ahead of the current time, at the county-level and weekly aggregated, in the United States. A notable feature of our spatio-temporal model is that it considers the (a) temporal auto- and pairwise correlation of the two local time series (confirmed cases and death of the COVID-19), (b) dynamics between locations (propagation between counties), and (c) covariates such as local within-community mobility and social demographic factors. The within-community mobility and demographic factors, such as total population and the proportion of the elderly, are included as important predictors since they are hypothesized to be important in determining the dynamics of COVID-19.
To reduce the model's high-dimensionality, we impose sparsity structures as constraints and emphasize the impact of the top ten metropolitan areas in the nation, which we refer (and treat within our models) as \emph{hubs} in spreading the disease. Our retrospective out-of-sample county-level predictions were able to forecast the subsequently observed COVID-19 activity accurately. 
The proposed multi-variate predictive models were designed to be highly interpretable, with clear identification and quantification of the most important factors that determine the dynamics of COVID-19. 
Ongoing work involves incorporating more covariates, such as education and income, to improve prediction accuracy and model interpretability. 
\end{abstract}

\section{Introduction}

The global spread of COVID-19, the disease caused by the novel coronavirus SARS-CoV-2, has affected nearly everyone's lives on the planet. Even the largest economies' resources have been strained due to the large infectivity and transmissibility of COVID-19. As the number of cases of COVID-19 continues increasing, understanding finer-grained spatio-temporal dynamics of this disease and some of the leading factors affecting disease transmissions is critical to helping officials make policy decisions and curb the further spread of the disease. 

Most of the previous research aimed at studying the spread of COVID-19 has focused on two key measurements: the number of confirmed cases and the number of deaths. Cases going up or down over time shed light on the rate of spread of COVID-19 at a given point in time — but it is only valid if enough people get tested. The limited testing ability resulted in a severe underestimation of COVID-19 cases in the pandemic's early stages \cite{kou2020unmasking}.     
For example, when there was not enough testing capacity, as was the case in New York City in March 2020, the number of cases reported was an undercount of actual cases, estimated to be much larger (up by a factor of 10) \cite{havers2020seroprevalence, lu2020estimating}. Some studies have circumvented underestimation by considering the case positivity rate, which measures the percentage of total COVID-19 tests conducted that are positive. 
However, most of the widely-used COVID-19 data sets, such as \emph{the COVID Tracking Project} \cite{Covidproj2020}, only collect the total number of people with a completed polymerize chain reaction (PCR) test that returns positive as reported by the state or territory, which has a much lower spatial resolution (state-level) in comparison with the cases and deaths data (county-level). 
Such coarse-grained testing numbers would introduce extra noise to our model and would most likely be incapable of improving the confirmed case prediction accuracy at the county level. 
Deaths are also an important metric that most people care about regarding the virus's ultimate epidemiological impact. 
In contrast to the number of confirmed cases, the number of deaths is a good and accurate indicator for evaluating how serious a burden this pandemic is causing, not only on health care systems but also on the general public's mental health and well-being. 
Some epidemiological studies, such as \cite{killian2020evaluating}, also recommend tracking deaths, even though deaths lag behind new cases, typically by two weeks to a month. 

\begin{figure}[b!]
\vspace{-.1in}
\centering
\includegraphics[width=.8\textwidth]{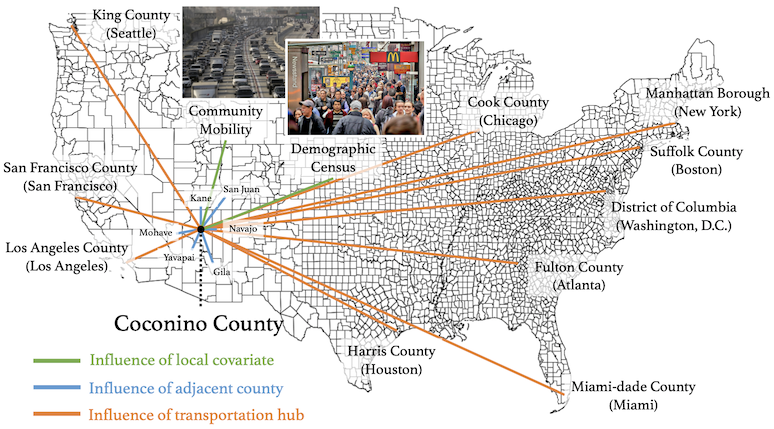}
\caption{An example of spatio-temporal covariates in our model for Coconino County, Arizona. Based on the counties in the U.S. as fundamental units, we assume the number of confirmed cases and deaths of COVID-19 reported in a given county are jointly related to the numbers reported in its adjacent counties (they are Kane, San Juan, Navajo, Gila, Yavapai, Mohave for Coconino in this example) and ten selected nationwide hubs (including San Francisco, Los Angeles, Seattle, Chicago, Atlanta, Miami, Washington, D.C., Boston, and New York). The numbers of cases and deaths also depend on some local covariates, such as community mobility level and some counties' demographic factors.}
\label{fig:model-exp}
\vspace{-.1in}
\end{figure}

A large amount of fine-grained data offers a unique opportunity to study the disease's spread dynamics from a micro-level view. 
For the United States, several teams have been working on collecting comprehensive COVID-19 tracking data, including daily counts of cases and deaths at the county level.
Such data gives us a general picture of how the virus is spreading across metropolitan and micropolitan counties and how such dynamics are evolving. Besides considering the cases and the deaths, we also aim to study other critical local factors in transmitting the COVID-19.
Recent studies \cite{Liu2020} on the spread of COVID-19 show that besides the distance to the epicenter, other factors, such as subway and airport, are positively connected with the virus transmission. 
Moreover, both urban areas and population density are positively associated with the spread of COVID-19 after the outbreak. 
The proportion of the elderly population has also been identified as a key factor in the death rate. Therefore, we consider the within-community mobility and two critical demographic factors by taking advantage of the COVID-19 Community Mobility Reports \cite{Covidcommunity2020} and the American Community Survey (ACS) \cite{ACS2019}. These two data sets are publicly available and include detailed county-level statistics that provide insights into what has changed in response to policies aimed at combating the COVID-19 and what factors may affect the disease's transmission. As illustrated in Fig.~\ref{fig:model-exp}, in our model, we assume the numbers of cases and deaths in each county depend on the neighboring counties and major metropolitan areas in the U.S., which we refer to as \emph{hubs} in spreading the disease. 
Local community mobility and demographic factors, including population and elderly population, are considered local covariates in the model, which also play a crucial role in the final number of deaths.

In this paper, we use a data-driven method incorporating a large-scale data set from multiple sources to predict the deaths and the confirmed cases of COVID-19 at the county level in the United States. 
Since death is a more accurate indicator for assessing the spread of the virus, we emphasize predicting county-level deaths’ trajectories instead of the confirmed cases. Our method's most notable contribution is considering the spatial structure among hubs and neighboring counties in modeling the cross-correlation between cases and deaths.  We also present the effect of a wide variety of geographic community mobility and social demographic factors on the spread of COVID-19. 
Our approach drastically differs from previous studies \cite{Colizza2006, Balcan2009, Adda2016}, in which the number of cases and deaths, and other covariates, including the community mobility and social demographic factors, are interlinked through a vector autoregressive process.  Our model shows that these hubs play a pivotal role in spreading the disease. We also find that both cases and deaths are significantly related to the local level's total population and that deaths are also positively associated with the proportion of the elderly population. Additionally, we found that confirmed cases are not significantly related to the proportion of the elderly population, which may prove that the disease was mostly circulating among young people in its later stage. In particular, while we identify a spike in cases since the beginning of the summer, we do not observe a clear spike in deaths. This may be explained by the fact that a more significant proportion of young people, who are generally at lower risk of death, were infected in the more recent pandemic stages.

The remainder of the article is organized as follows. We first review related works in the rest of this section, followed by describing the data sets we have used in Section~\ref{sec:data}. Section~\ref{sec:method} presents our proposed vector autoregressive model with spatial structure incorporated. 
We demonstrate the effectiveness of our model and discuss its interpretation in Section~\ref{sec:results}.
Lastly, the article concludes with discussions and future research directions in Section~\ref{sec:discussion}.

\paragraph{Related work} 

Compartmental models have been widely used in infectious disease epidemiological studies. In the SIR model \cite{Harko2014}, one of the simplest compartmental models, the population is assigned to three components: S (susceptible), I (infectious), and R (recovered). 
These variables (S, I, and R) represent the number of people in each compartment. 
The transition between different compartments is modeled using a set of coupled differential equations. 
Based on the SIR model, many variants have been proposed in the last decades, including the SIRD (Susceptible-Infectious-Recovered-Deceased) model \cite{fernandez2020estimating,caccavo2020chinese} that considers deceased individuals, and the
SEIR (Susceptible-Exposed-Infectious-Recovered) model \cite{Hethcote2000,yang2020modified,hou2020effectiveness,IHME,MOBS} that considers exposed period during which individuals have been infected but are not yet infectious themselves, to name a few. 
The total population is usually assumed to be fixed in the compartmental models; therefore, it works well when modeling nationwide data. However, in our high-resolution modeling, each county's population is of high variability due to dynamics across the county. Therefore, we use a spatio-temporal model instead to capture the influence of major big cities and neighboring counties without fixing each county's population. 

Besides compartmental models, much work has been done on predicting the total number of COVID-19 cases, and deaths without considering the spatial correlation across regions \cite{LANL,UTaustin,woody2020projections}. 
For example, 
recent work \cite{bertozzi2020challenges} introduces a regional model based on a self-exciting point process to forecast the total number of infections for multiple countries. 
Another work \cite{ghosh2020covid} provides a state-wise analysis and infections prediction for India's states by considering three growth models, namely, the logistic, the exponential, and the susceptible-infectious-susceptible models.
Machine-learning-based approaches have also been considered in modeling COVID-19 outbreak \cite{alazab2020covid}.
Some work \cite{tamang2020forecasting} attempts to use a neural network to model accumulative case counts for multiple countries.
Recurrent neural network-based methods \cite{hawas2020generated, zhao2020well} have been applied to model the temporal dynamics of the COVID-19 outbreak. Our approach differs from these studies in two ways: (1) our model provides finer-grained predictions for the cases and deaths; (2) we model the multivariate time series by considering the spatial correlation across regions as well as the correlation with the demographic factors, which is more interpretable than the machine-learning-based methods.

Understanding the COVID-19 outbreak's spatial spread is critical to predicting local outbreaks and developing public health policies during the early stages of COVID-19.
However, studies evaluating the spatial spread of the COVID-19 pandemic are scarce or limited \cite{poirier2020real}. 
Previous studies have described the spatial spread of severe acute respiratory syndrome (SARS) in Beijing and mainland China \cite{Meng2005, Fang2009, Kang2020, Jia2020, CCDCP2020} using limited or localized data. One study also considered the various connections between a few cities to calculate the spatial association \cite{Meng2005}. There is also prior work using the multivariate Hawkes process to model the conditional intensity of new COVID-19 cases and deaths in the U.S. at the county level \cite{chiang2020hawkes}, without considering the influence of the big cities and other important demographic factors. 
The work of \cite{altieri2020curating} develops two types of county-level predictive models based on the exponential and the linear model, respectively. It focuses on modeling the dynamics of cumulative death counts.
In \cite{kapoor2020examining}, the graph neural networks are adopted to capture the spatio-temporal dynamics between various features; However, the lack of interpretability hinders from further understanding the mechanism of the COVID-19 outbreak.

There is also a wide array of previous research based on autoregressive models that relate to our work. 
In \cite{agosto2020poisson, alzahrani2020forecasting, fadly2020approach, kirbacs2020comparative, roy2020spatial, singh2020prediction}, the autoregressive integrated moving average (ARIMA) is used to predict future data for different countries. 
\cite{maleki2020time} uses an autoregressive-based time series model to predict the total number of the world's confirmed cases. 
\cite{saba2020forecasting, chakraborty2020theta} adopt the autoregressive artificial neural networks to predict the number of accumulative cases in Egypt. 
A more recent similar article \cite{khan2020modelling} studies the state-wise cases in Pakistan using the vector autoregressive model.
However, two significant differences are (1) the spatial resolution of their predictions is much lower than our results; (2) these models do not consider the spatial correlation between places in the vicinity and cities served as major transportation hubs.

There are also various efforts studying impacts from other aspects, such as temperature, humidity \cite{Sajadi2020, poirier2020role}, age, gender \cite{CCDCP2020}, and travel restrictions \cite{Chinazzi2020, lai2020effect}.
Here, in our case, $N = I \times T = 157,200$. 
Most of these studies are constrained on a relatively small scale because of limited data at the pandemic's early stage.

\section{Data}
\label{sec:data}

\begin{figure}[!t]
\centering
\begin{subfigure}[h]{.31\linewidth}
\includegraphics[width=\linewidth]{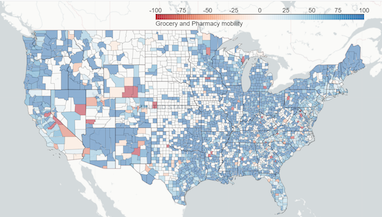}
\caption{grocery on Mar 1st}
\end{subfigure}
\begin{subfigure}[h]{.31\linewidth}
\includegraphics[width=\linewidth]{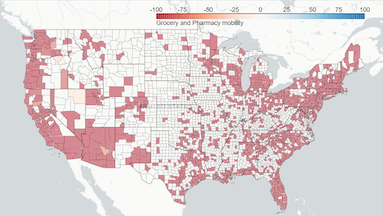}
\caption{grocery on April 12th}
\end{subfigure}
\begin{subfigure}[h]{.31\linewidth}
\includegraphics[width=\linewidth]{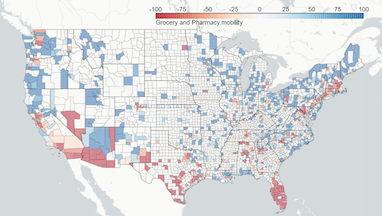}
\caption{grocery on July 12th}
\end{subfigure}
\vfill
\begin{subfigure}[h]{.31\linewidth}
\includegraphics[width=\linewidth]{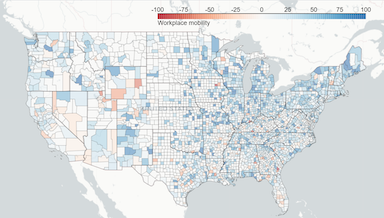}
\caption{workplace on Mar 1st}
\end{subfigure}
\begin{subfigure}[h]{.31\linewidth}
\includegraphics[width=\linewidth]{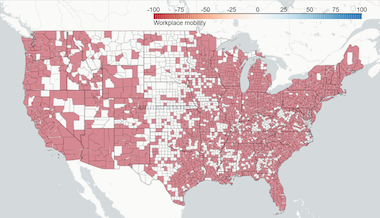}
\caption{workplace on April 12th}
\end{subfigure}
\begin{subfigure}[h]{.31\linewidth}
\includegraphics[width=\linewidth]{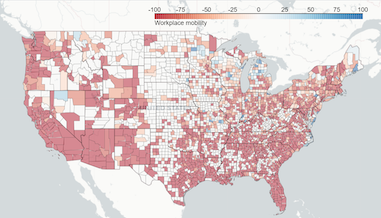}
\caption{workplace on July 12th}
\end{subfigure}
\vfill
\begin{subfigure}[h]{.31\linewidth}
\includegraphics[width=\linewidth]{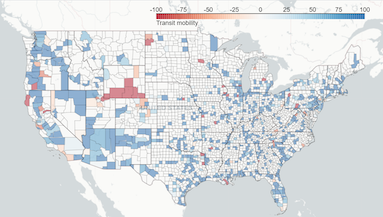}
\caption{transit on Mar 1st}
\end{subfigure}
\begin{subfigure}[h]{.31\linewidth}
\includegraphics[width=\linewidth]{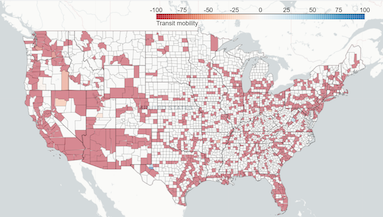}
\caption{transit on April 12th}
\end{subfigure}
\begin{subfigure}[h]{.31\linewidth}
\includegraphics[width=\linewidth]{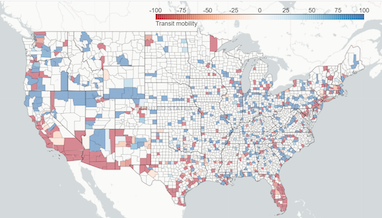}
\caption{transit on July 12th}
\end{subfigure}
\caption{Overview of Google mobility data in three selected categories: grocery, workplace, and transit on three different days. Counties in red and blue indicate their mobility is lower and higher than the normal level, respectively. The mobility level varies over time and space due to local government policy change in response to COVID-19.}
\label{fig:mobility-data}
\vspace{-.1in}
\end{figure}

We have used three comprehensive datasets in this study, including confirmed cases and deaths of COVID-19, community mobility data, and demographic census data. These datasets play an important role in understanding the spatio-temporal correlation of COVID-19 transmission. 

\paragraph{Confirmed cases and deaths of COVID-19}

We used the dataset from The New York Times \cite{NYT2019}, based on state and local health agencies' reports. 
The data is the product of dozens of journalists working across several time zones to monitor news conferences, analyze data releases, and seek public officials' clarification on how they categorize cases. 
The data includes two parts: (i) \emph{confirmed cases} are counts of individuals whose coronavirus infections were confirmed by a laboratory test and reported by a federal, state, territorial, or local government agency. Only tests that detect viral RNA in a sample are considered confirmatory. These are often called molecular or reverse transcription-polymerase chain reaction (RT-PCR) tests; (ii) \emph{confirmed deaths} are individuals who have died and meet the definition for a confirmed COVID-19 case. Some states reconcile these records with death certificates to remove deaths from their count, where COVID-19 is not listed as the cause of death. These data have removed non-COVID-19 deaths among confirmed cases according to the information released by health departments, i.e., in homicide, suicide, car crashes, or drug overdose.
All cases and deaths are counted on the date they are first announced.
In practice, we have observed periodic weekly oscillations in daily reported cases and deaths, which could have been caused by testing bias (higher testing rates on certain days of the week). 
To reduce such bias, we aggregate the number of cases and deaths of each county {\it by week}.

\paragraph{Community mobility} 

As global communities respond to COVID-19, we have heard from public health officials that the same type of aggregated, anonymized insights we use in products such as Google Maps could be helpful as they make critical decisions to combat COVID-19. The COVID-19 Community Mobility Reports \cite{Covidcommunity2020} aim to provide insights into what has changed in response to policies aimed at combating COVID-19.  The reports record people's movement by county daily, across various categories such as retail and recreation, groceries and pharmacies, parks, transit stations, workplaces, and residential. The data shows how visitors to (or time spent in) categorized places change compared to the baseline days (in percentage). The negative percentage represents the level of mobility is lower than the baseline, and the positive percentage represents the opposite. 
A baseline day represents a normal value for that day of the week. The baseline day is the median value from the five weeks from January 3rd to February 6th, 2020. To match the temporal resolution with the COVID-19 data and detrend the weekly pattern, we aggregate each county's mobility data by week. 
Examples of three categories have been shown in Fig.~\ref{fig:mobility-data}.

\begin{figure}[!t]
\centering
\begin{subfigure}[h]{.49\linewidth}
\includegraphics[width=\textwidth]{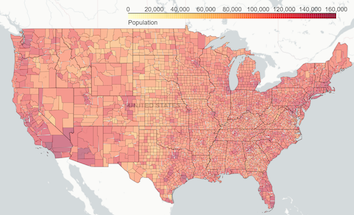}
\caption{county-level total population}
\end{subfigure}
\begin{subfigure}[h]{.49\linewidth}
\includegraphics[width=\textwidth]{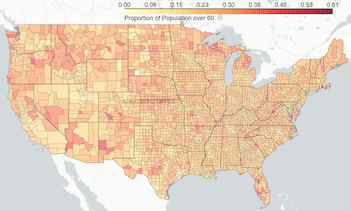}
\caption{county-level proportion of elderly population}
\end{subfigure}
\caption{Overview of the social demographic factors. The color depth represents the value of the demographic variables of interest in certain counties.}
\label{fig:census-data}
\vspace{-.1in}
\end{figure}

\paragraph{Demographic census} 

Data from the American Community Survey (ACS) \cite{ACS2019},  provided by the U.S. Census Bureau, is a comprehensive source for demographic information about the population, age, and economic status in each zip code region in the U.S. 
Unlike the census data, which takes place every ten years, the ACS is conducted every year. The latest ACS data are available in the year 2018. Some demographic factors help us understand how population distribution affects the spread of disease (by correlating the local socio-economic profile with its confirmed cases and deaths). These factors contain essential information about the development and economic growth of different areas. To match the spatial resolution with the COVID-19 data, we aggregate the zip code regions' demographic data in the same county. 
We selected two leading factors that affect the spread and the infection of the disease, i.e., total population and the proportion of the elderly with an age of 65 or older \cite{onder2020case}.

\section{Methodology}
\label{sec:method}

This section presents our statistical model that captures the spatio-temporal correlation of the spread of COVID-19. We begin with a brief description of the problem setup and notations, then jointly model confirmed cases and deaths as a vector autoregressive process in Section \ref{sec:spatio-temporal-model}. The essential notations defined in this section are also summarized in Table~\ref{tab:notation}. 

\begin{table}[!ht]
\caption{Summary of essential notations}
\label{tab:notation}
\vspace{-0.1in}
\resizebox{\textwidth}{!}{%
\begin{tabular}{lll}
\toprule[1pt]\midrule[0.3pt]
\bf Section & \bf Notation & \bf Description \\ \hline
\ref{sec:prob-setup}
& $\mathcal{T} = \{t = 1, \dots, T\}$ & Set of all weeks.\\
& $\mathcal{I} = \{i = 1,\dots,N\}$ & Set of all counties.\\
& $\mathcal{K} = \{k = 1,\dots,K\}$ & Set of mobility categories.\\
& $\mathcal{L} = \{k = 1,\dots,L\}$ & Set of demographic factors.\\
& $c_{i, t} \in \mathbb{Z}_+$ & Number of confirmed cases for county $i \in \mathcal{I}$ in week $t \in \mathcal{T}$.\\ 
& $d_{i, t} \in \mathbb{Z}_+$ & Number of deaths for county $i \in \mathcal{I}$ in week $t \in \mathcal{T}$.\\
& $z_{i,l} \in \mathbb{R}_+$ & Data of demographic factor $l \in \mathcal{L}$ for county $i \in \mathcal{I}$.\\
& $m_{i,k,t} \in \mathbb{R}$ & Data of mobility category $k \in \mathcal{K}$ for county $i \in \mathcal{I}$ in week $t \in \mathcal{T}$.\\
\hline
\ref{sec:spatio-temporal-model} 
& $\mathcal{A} = \{(i, j):~i, j\in\mathcal{I}\}$ & Set of all county pairs in the U.S. that $i, j$ are adjacent to each other or one of $i,j$ is a \emph{hub}.\\
& $\boldsymbol B_\tau = (\beta_{i, j}) \in \mathbb{R}^{N \times N}$ & Case's coefficients depended on past confirmed cases between county $i, j \in \mathcal{I}$ for $\tau$ weeks ago.\\
& $\boldsymbol A_\tau = (\alpha_{i, j}) \in \mathbb{R}^{N \times N}$ & Death's coefficients depended on past deaths between county $i, j \in \mathcal{I}$ for $\tau$ weeks ago.\\
& $\boldsymbol H_\tau = (h_{i, j}) \in \mathbb{R}^{N \times N}$ & Death's coefficients depended on past confirmed cases between county $i, j \in \mathcal{I}$ for $\tau$ weeks ago.\\
& $\mu_{k, \tau} \in \mathbb{R}$ & Coefficient for mobility category $k \in \mathcal{K}$ in the past $\tau$-th week 
w.r.t. the number of cases.\\
& $\nu_{k, \tau} \in \mathbb{R}$ & Coefficient for mobility category $k \in \mathcal{K}$ in the past $\tau$-th week
w.r.t. the number of deaths.\\
& $\upsilon_l \in \mathbb{R}$ & Coefficient for demographic factor $l \in \mathcal{L}$ 
w.r.t. the number of cases. \\
& $\zeta_l \in \mathbb{R}$ & Coefficient for demographic factor $l \in \mathcal{L}$
w.r.t. the number of deaths. \\
\midrule[0.3pt]\bottomrule[1pt]
\end{tabular}%
}
\end{table}

\subsection{Problem setup and notations} 
\label{sec:prob-setup}

Consider confirmed cases and deaths of the COVID-19 in $N$ counties and $T$ weeks (recall that we aggregated these numbers by week to reduce bias). 
Let $\mathcal{I} = \{i=1, \dots, N\}$ be the set of counties and $\mathcal{T} = \{t=1, \dots, T\}$ be the set of weeks starting from March 15th, 2020 until January 17th, 2021. 
We assume there is a set of counties $\mathcal{I}' = \{i=1,\dots,N'\} \subset \mathcal{I}$ playing a significant role in spreading the disease due to their high population density and well-developed transportation network connecting to other major cities in the U.S..
We refer to these counties as \emph{hubs}, and the selected hubs are marked in Fig.~\ref{fig:model-exp}. 
Denote the number of confirmed cases and deaths in county $i \in \mathcal{I}$ and week $t \in \mathcal{T}$ as $c_{i,t} \in \mathbb{Z}_+$ and $d_{i,t} \in \mathbb{Z}_+$, respectively. In our setting, $T = 49$, $N = 3144$, and $N' = 10$.

We also consider $K$ mobility categories and $L$ demographic factors as covariates of the model, where $K = 6$ and $L = 2$. 
Let $\mathcal{K} = \{k=1,\dots,K\}$ be the set of community mobility categories and $\mathcal{L} = \{l=1,\dots,L\}$ be the set of demographic factors. 
Denote the mobility score in category $k \in \mathcal{K}$ for county $i \in \mathcal{I}$ in week $t \in \mathcal{T}$ as $m_{i,k,t} \in \mathbb{R}$, and denote the data of demographic factor $l \in \mathcal{L}$ for county $i \in \mathcal{I}$ as $z_{i,l} \in \mathbb{R}_+$. Let $\boldsymbol{c}_t \coloneqq [c_{1,t}, \dots, c_{N,t}]^\intercal$ and $\boldsymbol{d}_t \coloneqq [d_{1,t}, \dots, d_{N,t}]^\intercal$ denote the confirmed cases and deaths in week $t \in \mathcal{T}$, respectively.
Let $\boldsymbol{m}_{k,t} \coloneqq [m_{1,k,t}, \dots, m_{N,k,t}]^\intercal$ denote the score of community mobility category $k \in \mathcal{K}$ for all counties $i \in \mathcal{I}$ in week $t \in \mathcal{T}$. 
Let $\boldsymbol{z}_{l} \coloneqq [z_{1,l},\dots,z_{N,l}]^\intercal$ denote the data of demographic factor $l \in \mathcal{L}$ for all counties $i \in \mathcal{I}$.

\subsection{Spatio-temporal vector autoregressive model}
\label{sec:spatio-temporal-model}

\begin{figure}[t!]
\vspace{-.1in}
\centering
\includegraphics[width=.8\textwidth]{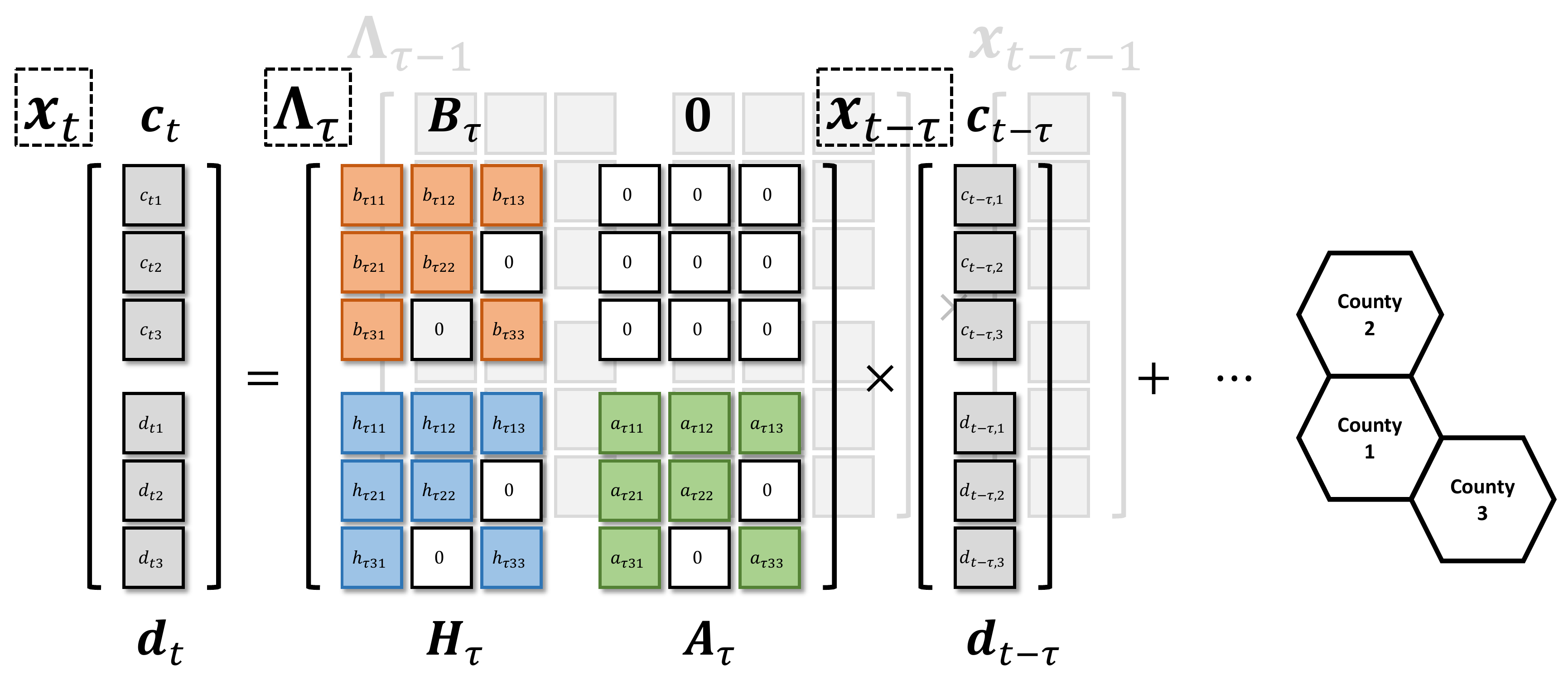}
\caption{A small illustrative example of spatial factor matrices with three counties. The adjacency of these counties is shown on the right. In this example, the observation $x_t$ is 6-dimensional and the matrices $\boldsymbol A_\tau$, $\boldsymbol H_\tau$ and $\boldsymbol B_\tau$ are all 3-by-3 matrices. The white boxes represent zero entries; The gray boxes represent data entries (cases and deaths); the red, the blue, and the green boxes represent the learnable (non-zero) entries in matrix $\boldsymbol{B}_\tau, \boldsymbol{H}_\tau, \boldsymbol{A}_\tau$, respectively. }
\label{fig:matrix-exp}
\vspace{-.1in}
\end{figure}

We consider a linear spatio-temporal autoregressive model where the number of confirmed cases ($\boldsymbol{c}_t$) and deaths ($\boldsymbol{d}_t$) is a time series regressed on their previous values and the mobility covariate $\boldsymbol{m}_{k,t}$ and demographic covariate $\boldsymbol{z}_l$. Denote the time window's length that we consider in the past (the memory depth) as $p$. Based on previous studies \cite{singhal2020review}, it is known that the COVID-19 virus has an incubation period of around two weeks. Therefore, we choose $p=2$ throughout this paper.

Define the augmented observation vector as (which contains both confirmed case and death counts):  
\[
\boldsymbol{x}_t \coloneqq \begin{bmatrix}\boldsymbol{c}_t\\ \boldsymbol{d}_t
\end{bmatrix} \in \mathbb R ^{2N}.
\]

Then our spatio-temporal model can be written as a vector autoregressive (VAR) process:
\begin{equation}
\boldsymbol{x}_t =
\sum_{\tau=1}^p \boldsymbol \Lambda_\tau \boldsymbol x_{t-\tau} +
\sum_{k=1}^K \sum_{\tau=1}^p \boldsymbol \gamma_{k,\tau} \otimes \boldsymbol{m}_{k,t-\tau} + 
\sum_{l=1}^L \boldsymbol{\omega}_l \otimes \boldsymbol{z}_{l}+ 
\boldsymbol \epsilon_t, \quad
\boldsymbol \epsilon_t \sim \mathcal N\left(\boldsymbol 0, \begin{bmatrix}
\boldsymbol \Sigma_\eta & \boldsymbol 0 \\
\boldsymbol 0 & \boldsymbol \Sigma_\eta
\end{bmatrix}\right),
\label{eq:case-death-model}
\end{equation}
where $\otimes$ is the Kronecker product and 
\[
\boldsymbol \Lambda_\tau =
\begin{bmatrix}
\boldsymbol B_\tau & \boldsymbol 0\\
\boldsymbol H_\tau & \boldsymbol A_\tau
\end{bmatrix} \in \mathbb{R}^{2N \times 2N}, 
~
\boldsymbol{\gamma}_{k,\tau} = 
\begin{bmatrix}
\mu_{k,\tau} \\
\nu_{k,\tau} 
\end{bmatrix} \in \mathbb{R}^{2},
~
\boldsymbol{\omega}_l = 
\begin{bmatrix}
\upsilon_l \\
\zeta_l
\end{bmatrix} \in \mathbb{R}^{2},
~
\boldsymbol{\epsilon}_t = 
\begin{bmatrix}
\boldsymbol{\epsilon}_{t,c} \\
\boldsymbol{\epsilon}_{t,d}
\end{bmatrix} \in \mathbb{R}^{2 N},~
1 \le \tau \le p.
\]
In our model \eqref{eq:case-death-model}, the first-term captures the dependence on past confirmed cases and deaths; the second term captures the influence of past local community mobility; the third term captures the influence of local demography, which is held constant over time. 
Specifically, $\boldsymbol B_\tau$ and $\boldsymbol{A}_\tau$ contain the autoregressive coefficients for the number of confirmed cases and deaths, respectively; $\boldsymbol{H}_\tau$ describes the dependence of the current number of deaths on the number of confirmed cases in $\tau$ weeks ago. 
As an illustrative example shown in Fig.~\ref{fig:matrix-exp},
these three matrices share the same sparse structure, where the entry at $(i, j)$ is zero if county $i$ and county $j$ are not adjacent and none of them is the hub.
Formally, the set of adjacency pairs is defined by $\mathcal{A} = \{(i,j)\in \mathcal{I}: (i,j)\text{ is an edge of the graph }\mathcal{G}\}$; each node of $\mathcal{G}$ denotes a county, and there is an edge between two nodes whenever the corresponding counties are geographically adjacent or one of them is a hub.
The $\mu_{k,\tau}$, $\nu_{k,\tau}$, $\upsilon_l$, and $\zeta_l$ are four scalar coefficients. 
To be specific, $\mu_{k,\tau}$, $\nu_{k,\tau}$ represent the coefficients for the local community mobility score in category $k$ in $\tau$ weeks ago with respect to the corresponding number of confirmed cases and deaths, respectively. 
Similarly, $\upsilon_l$, $\zeta_l$ represent the coefficients for local demographic factor $l$ with respect to the corresponding number of confirmed cases and deaths, respectively.
The spatial covariance matrix between the noise at counties $i$ and $j$ is denoted as the $(i,j)$-th entry of $\boldsymbol \Sigma_\eta$; it is a function of their Euclidean distance $s_{ij}$ and is parameterized by $\eta$. Some commonly used spatial models include:
Gaussian model \cite{Lee2014}, Exponential model \cite{Gaetan2010}, and Mat\'{e}rn model \cite{Gaetan2010}.
Here we adopt the exponential spatial covariance model $\boldsymbol \Sigma_\eta(i, j) = \eta \exp \{ - \eta s_{ij} \}$, where $\eta$ is a pre-specified parameter, which controls the rate of spatial decay. In this paper, we specify a reasonable value of the parameter $\eta = 10^3$.

We aim to fit the model \eqref{eq:case-death-model} for confirmed cases and deaths jointly by minimizing the \emph{prediction error}. 
Define the set of parameters $\boldsymbol \theta = \{\boldsymbol{\Lambda}, \boldsymbol{\omega}, \boldsymbol{\gamma}\} \in \boldsymbol{\Theta}$, where $\boldsymbol \Theta$ is the set containing all feasible values. For a pre-specified hyper-parameter $\delta \in [0,1]$, the loss function is defined as a weighted combination of quadratic loss functions for death and confirmed case residuals: 
\begin{equation}
\ell(\boldsymbol \theta) \coloneqq \delta 
\sum_{t=1}^T \boldsymbol \varepsilon_{t, \rm d}^\intercal  \boldsymbol \Sigma_\eta^{-1} \boldsymbol \varepsilon_{t, \rm d}
+ (1-\delta) 
\sum_{t=1}^T \boldsymbol \varepsilon_{t, \rm c}^\intercal  \boldsymbol \Sigma_\eta^{-1} \boldsymbol \varepsilon_{t, \rm c},
\label{eq:obj}
\end{equation}
where $\boldsymbol \varepsilon_{t, \rm c}$ denotes the confirmed case prediction residual
\[
\boldsymbol \varepsilon_{t,\rm c} = \begin{bmatrix}
\boldsymbol I & \boldsymbol 0
\end{bmatrix}
\left (\boldsymbol x_t-
\sum_{\tau=1}^p \boldsymbol \Lambda_\tau \boldsymbol x_{t-\tau} -
\sum_{k=1}^K \sum_{\tau=1}^p \boldsymbol \gamma_{k, \tau} \otimes \boldsymbol{m}_{k,t-\tau} - 
\sum_{l=1}^L \boldsymbol{\omega}_l \otimes \boldsymbol{z}_{l}
\right ),
\]
and $\boldsymbol \varepsilon_{t, \rm d}$ denotes the death prediction residual 
\[
\boldsymbol \varepsilon_{t,\rm d} = \begin{bmatrix}
\boldsymbol 0 & \boldsymbol I
\end{bmatrix}
\left (\boldsymbol x_t-
\sum_{\tau=1}^p \boldsymbol \Lambda_\tau \boldsymbol x_{t-\tau} -
\sum_{k=1}^K \sum_{\tau=1}^p \boldsymbol \gamma_{k, \tau} \otimes \boldsymbol{m}_{k,t-\tau} - 
\sum_{l=1}^L \boldsymbol{\omega}_l \otimes \boldsymbol{z}_{l}
\right ).
\]
The hyper-parameter $\delta$ controls the proportion of death prediction loss. In practical terms, we emphasize the importance of death, and hence we choose $\delta = 0.9$ empirically. The reason is that it is known that the confirmed cases are quite noisy and can depend on the capacity of testings. 

The parameters $\boldsymbol \theta$ can be estimated by solving the following optimization with a regularization function: 
\begin{equation}
\min_{\boldsymbol \theta \in \boldsymbol \Theta} \ell(\boldsymbol \theta) + \lambda_1 R (\boldsymbol \theta ),
\label{eq:opt}
\end{equation}
where $\lambda_1\geq 0$ is a parameter that controls the importance of the regularization term, and $R(\boldsymbol\theta)$ is the elastic net type regularization function (with hyper-parameter $\lambda_2 \in [0,1]$) given by
\[
R(\boldsymbol \theta):= \sum_{\tau=1}^p  \sum_{i=1}^N \sum_{j=1}^N \mathds{1}_\mathcal{A} \left\{ (i, j) \right\} \left[
\lambda_2 \Big(|\alpha_{i,j,\tau}| + |\beta_{i,j,\tau}| + |h_{i,j,\tau}|\Big) +  (1-\lambda_2) \Big(|\alpha_{i,j,\tau}|^2 + |\beta_{i,j,\tau}|^2 + |h_{i,j,\tau}|^2\Big)
\right],
\]
where $\mathds{1}_A\{x\}$ is the indicator function, i.e., taking the value 1 if $x \in A$ otherwise 0; $\lambda_2$ is the $\ell_1$ penalty ratio in the regularization function; $\alpha_{i,j,\tau}, \beta_{i,j,\tau}, h_{i,j,\tau}$ are the entries of matrices $\boldsymbol A_\tau, \boldsymbol B_\tau, \boldsymbol H_\tau$, respectively. 


\subsection{Exploit sparsity and structure to solve large-scale optimization problems}

Our model's most salient feature is that we consider the underlying spatio-temporal structure between the number of confirmed cases and deaths. If there is no specific structure in coefficient matrices, our methods look on the surface to be a naive linear model but require to solve a large-scale high-dimensional optimization problem, which contains 79,077,916 parameters (variables in the optimization problem) with only 84,888 data points. 
Instead of solving such complex problems directly, we tackle this challenge by exploiting the sparse spatial structure and only consider the correlation between adjacent counties and hubs, which leads to a significant reduction in the number of parameters (less than 80,000). 
Besides, the lower triangular structure of $\boldsymbol \Lambda_\tau$ matrix (including $\boldsymbol B_\tau$, $\boldsymbol H_\tau$, and $\boldsymbol A_\tau$) captures the causal relationship we believe exists in the confirmed case to the death count, but not the other way around. To be exact, we assume the number of confirmed cases in the past will result in the change of both the confirmed cases and deaths in the future, while the number of deaths only relates to the future's deaths. 

The regularization term we devised in Section~\ref{sec:spatio-temporal-model} also plays a big part in achieving the ideal results. This elastic net-based method linearly combines the lasso and ridge regression penalties on $\boldsymbol B_\tau$, $\boldsymbol H_\tau$, and $\boldsymbol A_\tau$ to encourage sparse spatial correlation and stabilize the solution at the same time. The hyper-parameters $\lambda_1$ and $\lambda_2$ in the regularization term are chosen by $5$-fold cross-validation, where the optimal choices are $\lambda_1 \approx 10^2$ and $\lambda_2 \approx 10^{-1}$ for the fitted model. 

Here we solve the optimization problem by gradient descent. To fit the model, we first standardize the data of covariates and feed all the data as a single batch in one iteration, then descend the gradients of the parameters with respect to the loss defined in \eqref{eq:obj} until the model converges. To perform a one-week ahead prediction, we feed all the data before that week as a single batch in one iteration and follow the same gradient descent procedure described above. The model normally takes about 500 iterations to reach the convergence with $\ell(\boldsymbol \theta) \approx 1.41 \times 10^3$.

\section{Results}
\label{sec:results}

Now we report the results of our study. We evaluate the explanatory power of proposed modeling method by performing the in-sample estimation. We also compare our approach regarding the one-week ahead predictive performance against four other external benchmark methods, which are the current state-of-art methods adopted by the Centers for Disease Control and Prevention (CDC) in its national ensemble forecast \cite{CDC, ray2020ensemble}.
We generate the death (and case) prediction for each county in a week by taking the past county-level case and death records, community mobility, and demographic census information as input. The format of the input data  is described in Section~\ref{sec:data}.
In addition, we demonstrate the interpretable components of our model by showing the spatio-temporal correlation between the number of COVID-19 cases and other covariates discovered from our fitted model.
For the ease of presentation, we only focus on the mainland and do not consider Hawaii, Alaska, and other unincorporated territories of the United States in this paper.
Hereinafter, we refer to the proposed spatio-temporal vector autoregressive model as \texttt{STVA}.

\subsection{Model evaluation}

\begin{figure}[!t]
\centering
\begin{subfigure}[h]{.325\linewidth}
\includegraphics[width=\textwidth]{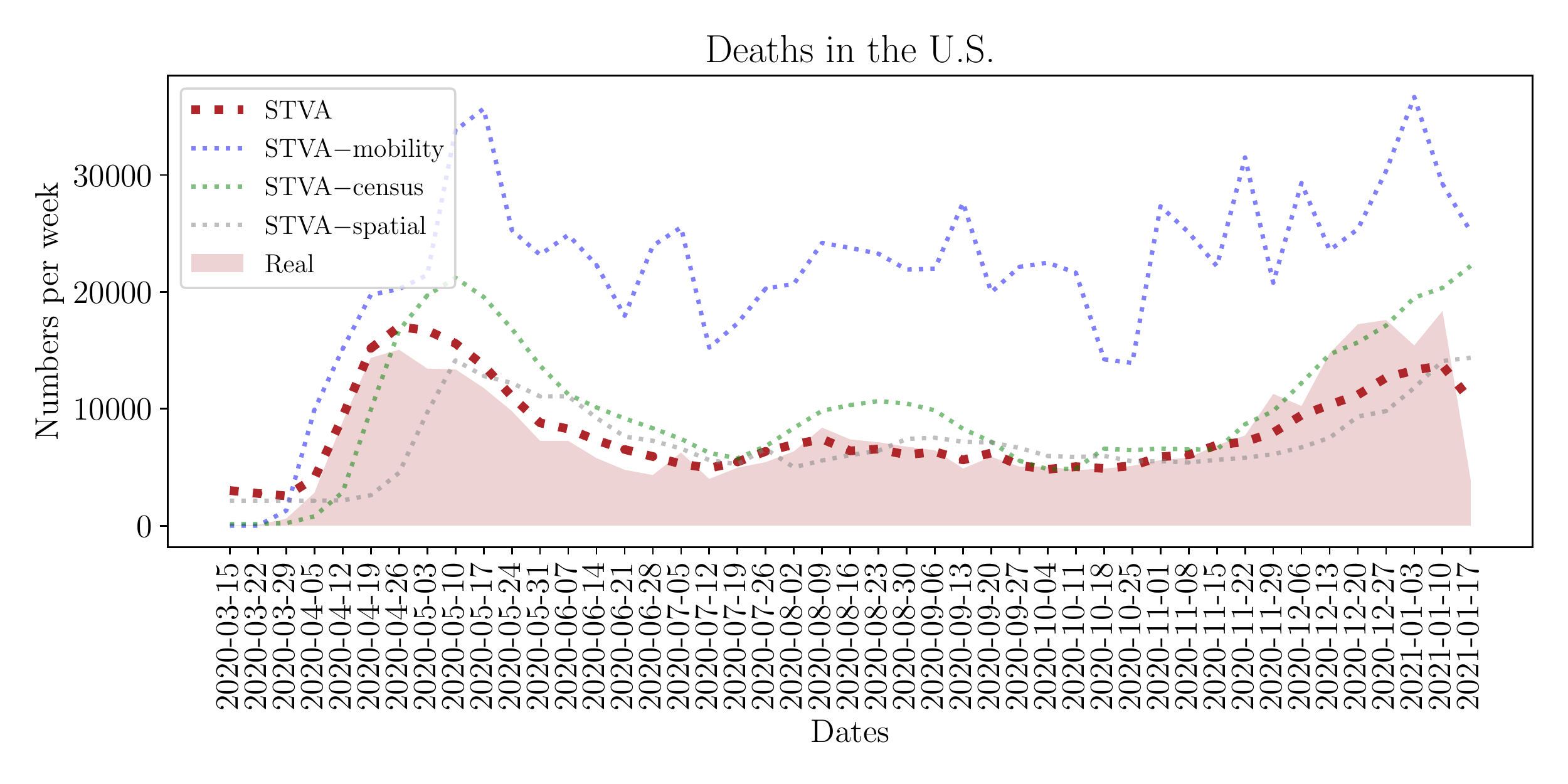}
\end{subfigure}
\begin{subfigure}[h]{.325\linewidth}
\includegraphics[width=\textwidth]{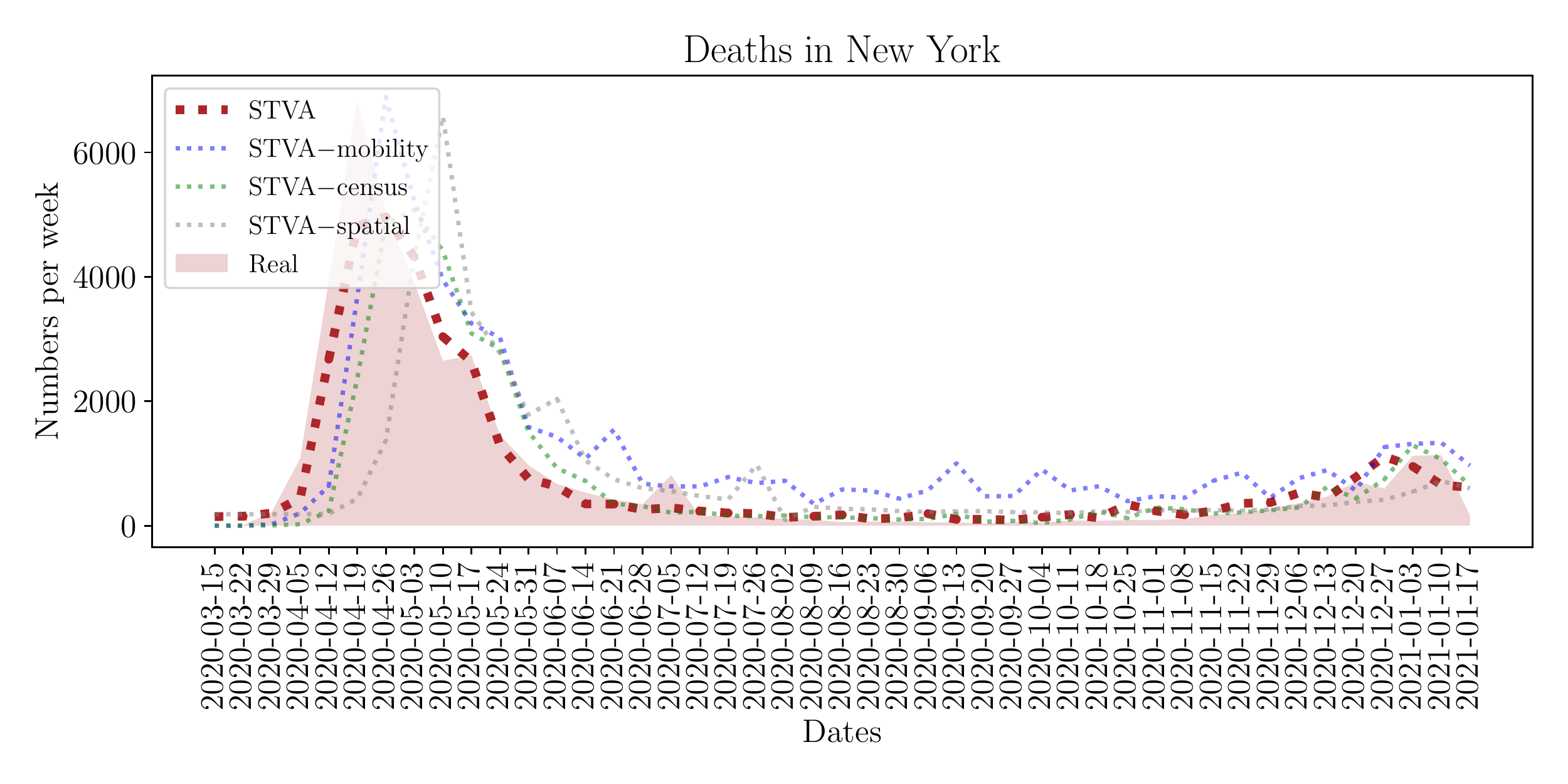}
\end{subfigure}
\begin{subfigure}[h]{.325\linewidth}
\includegraphics[width=\textwidth]{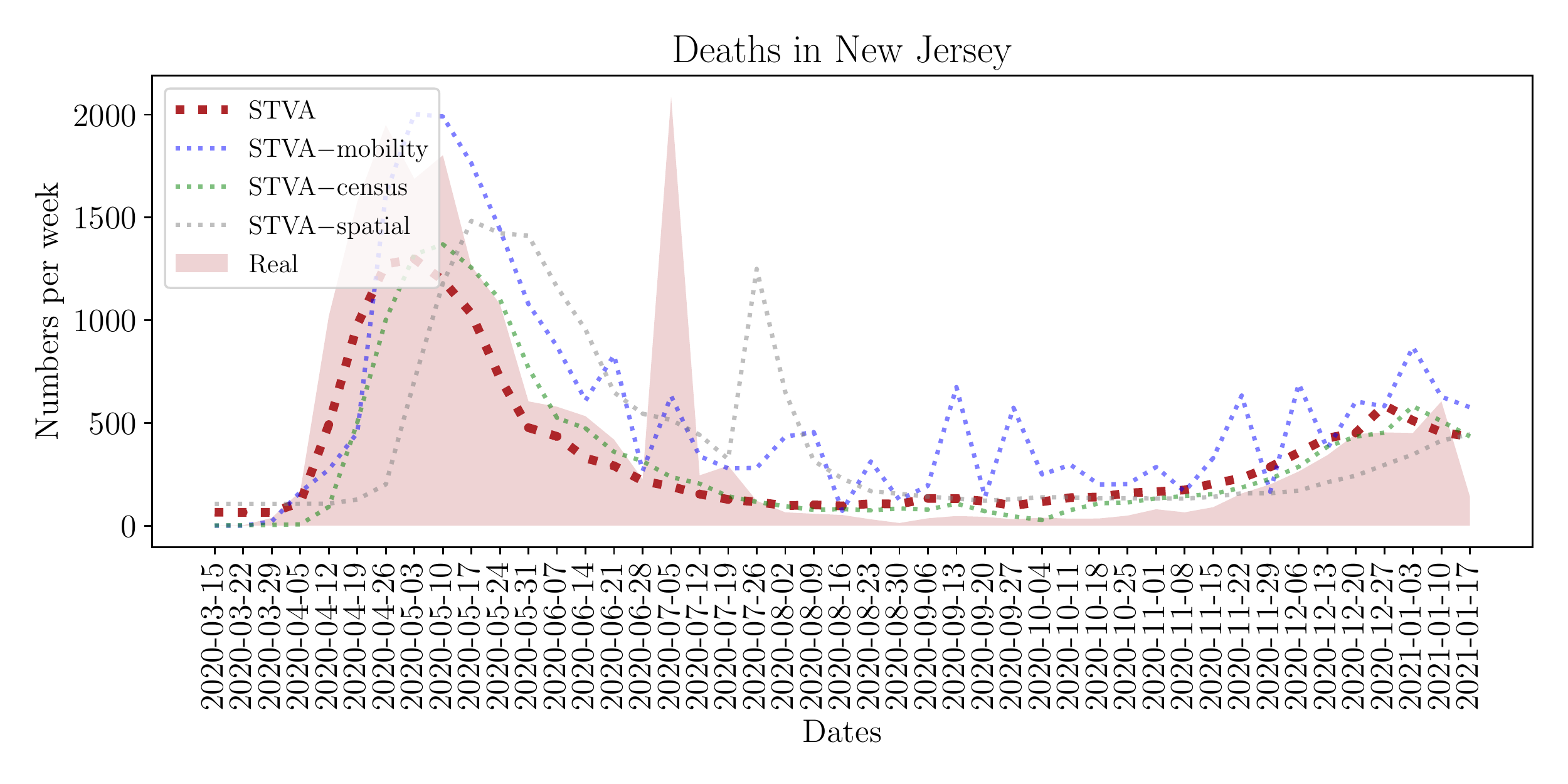}
\end{subfigure}
\vfill
\begin{subfigure}[h]{.325\linewidth}
\includegraphics[width=\textwidth]{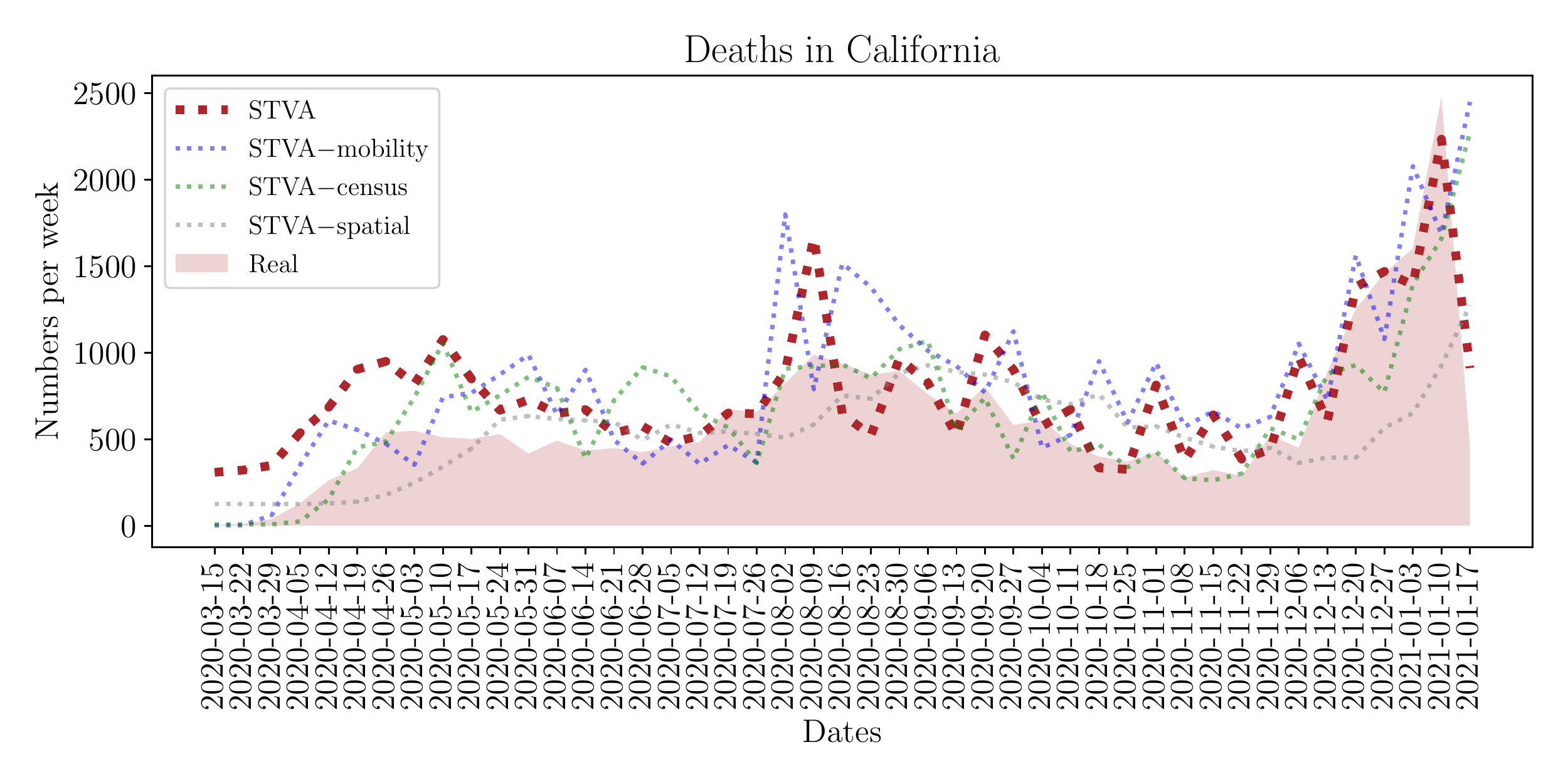}
\end{subfigure}
\begin{subfigure}[h]{.325\linewidth}
\includegraphics[width=\textwidth]{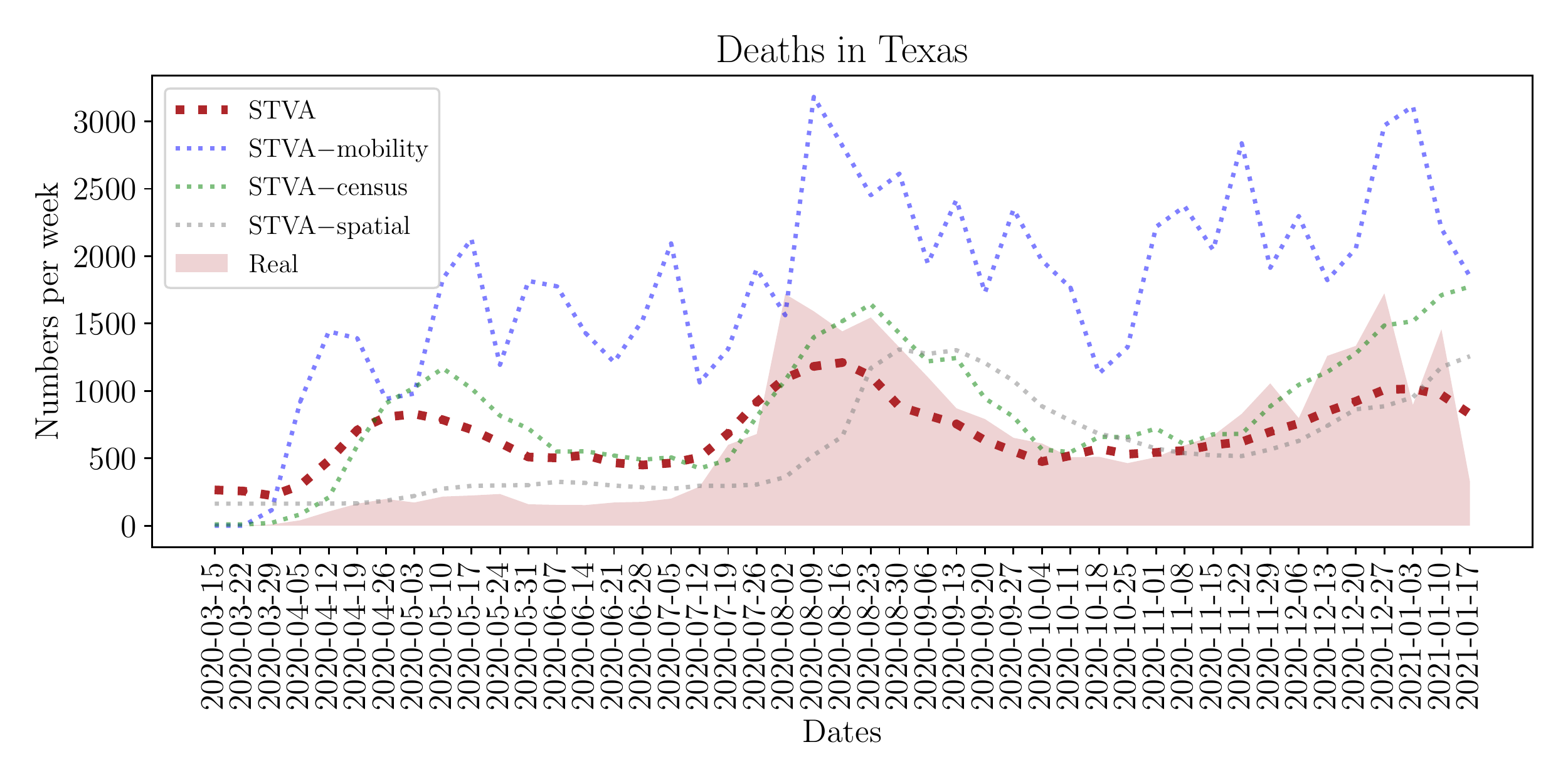}
\end{subfigure}
\begin{subfigure}[h]{.325\linewidth}
\includegraphics[width=\textwidth]{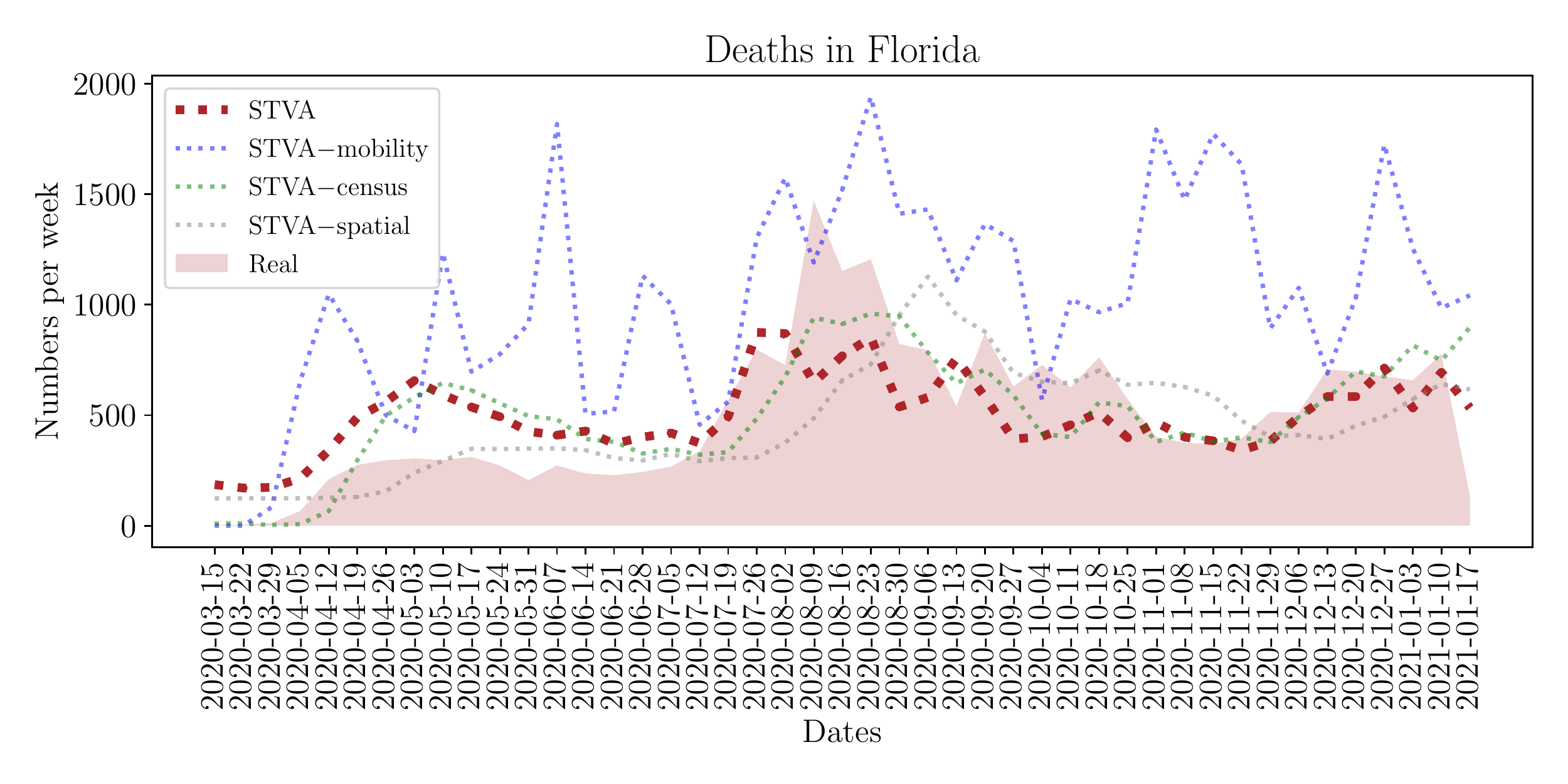}
\end{subfigure}
\vfill
\begin{subfigure}[h]{.325\linewidth}
\includegraphics[width=\textwidth]{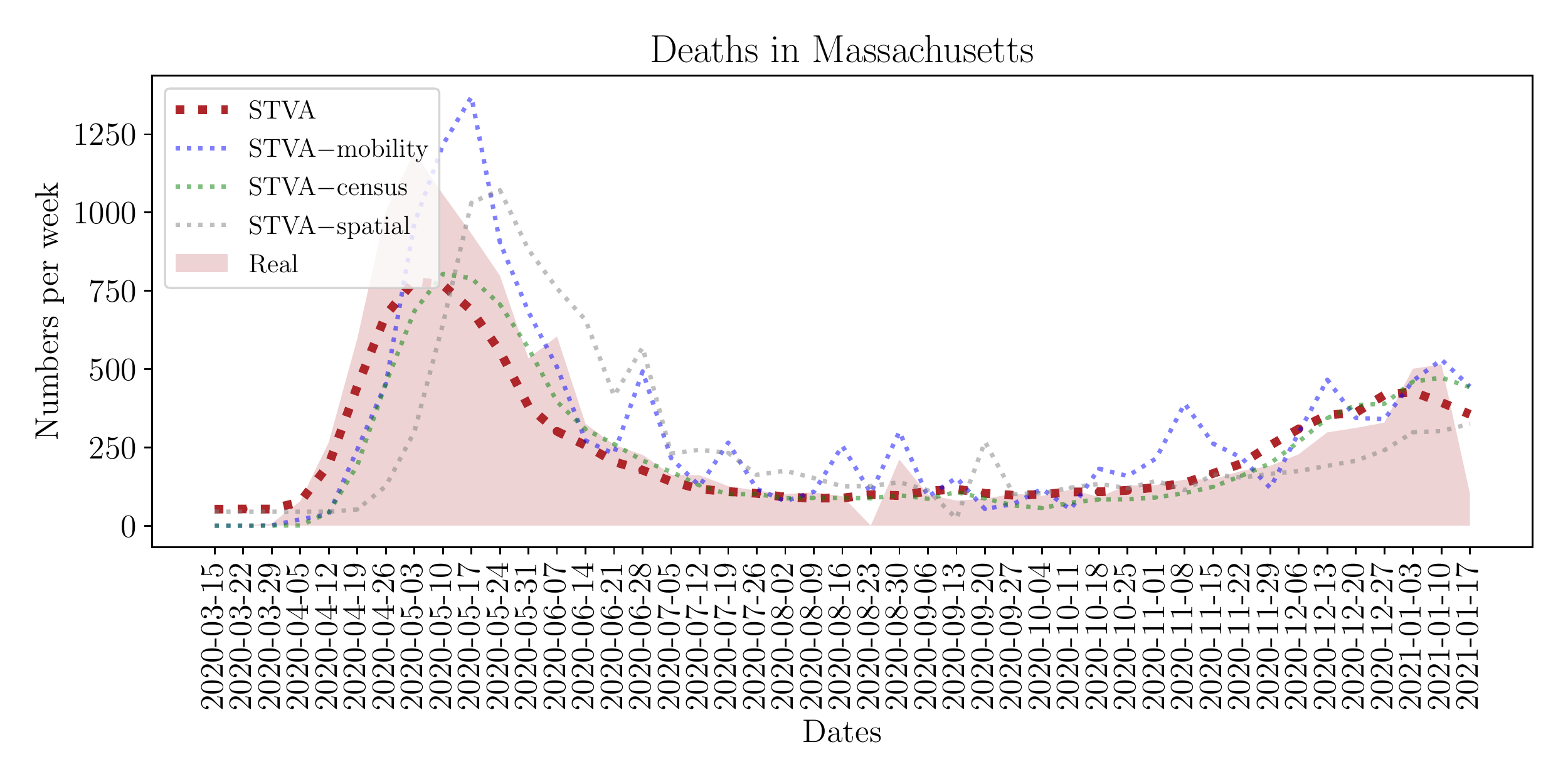}
\end{subfigure}
\begin{subfigure}[h]{.325\linewidth}
\includegraphics[width=\textwidth]{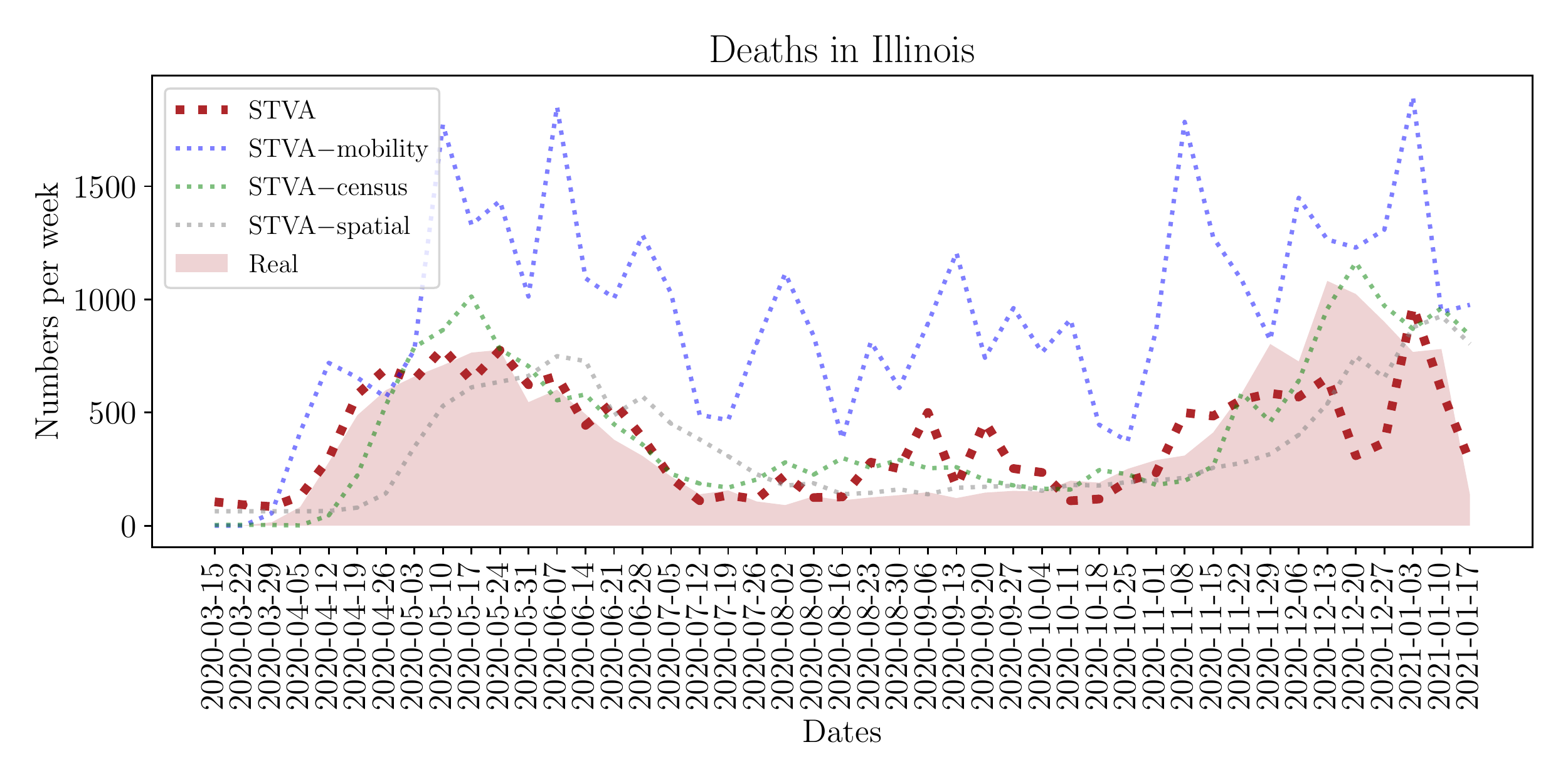}
\end{subfigure}
\begin{subfigure}[h]{.325\linewidth}
\includegraphics[width=\textwidth]{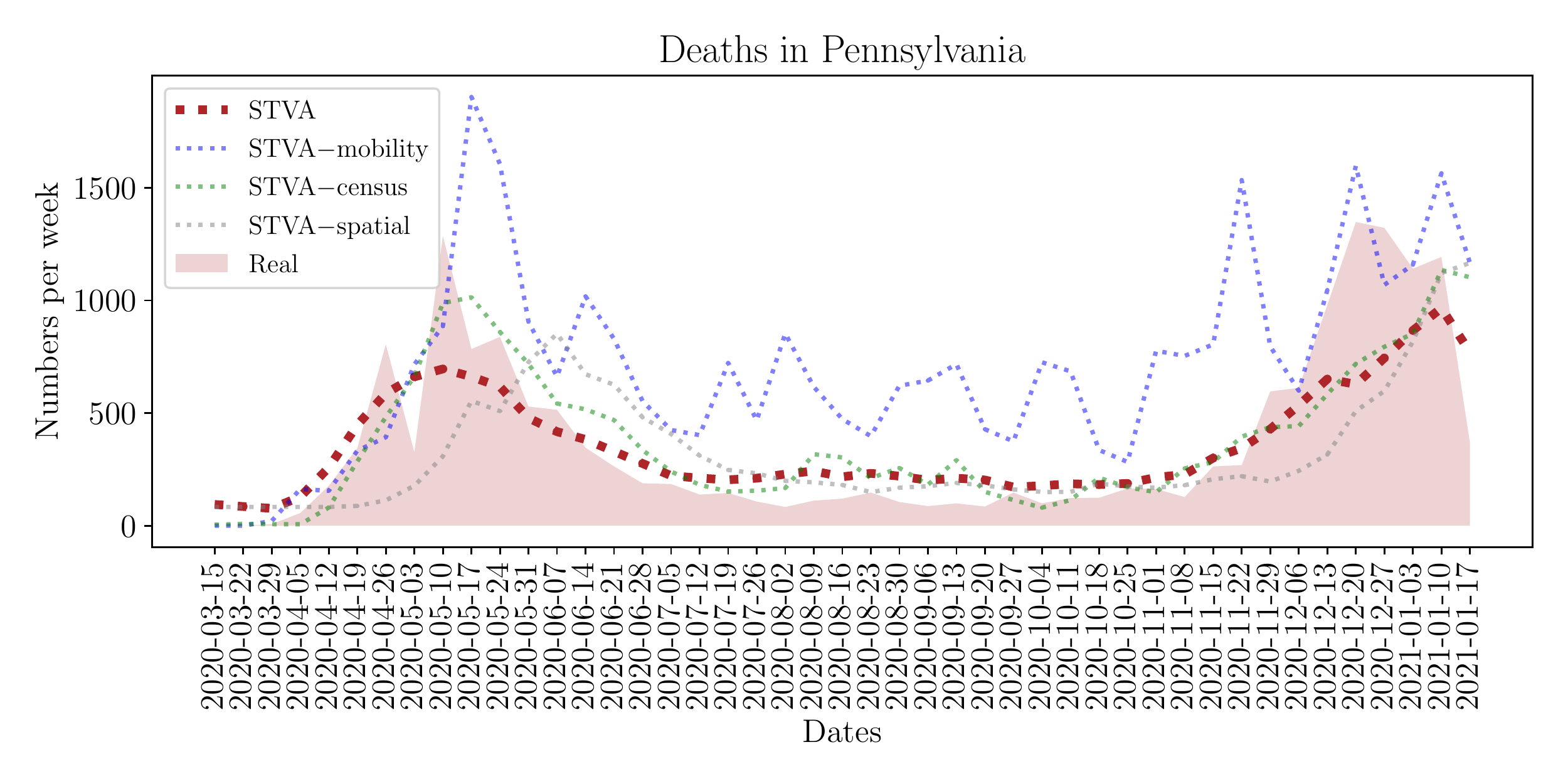}
\end{subfigure}
\caption{In-sample estimated deaths (red dotted lines) for the U.S. and other eight major states with the highest number of COVID-19 deaths in the U.S.. Figures are sorted in descending order of the total number of deaths since March 15th, 2020. The results show that our model can capture the general dynamics of the death numbers.}
\label{fig:statewise-death}
\vspace{-.1in}
\end{figure}


\begin{table}[]
\caption{Quantitative comparison of MAE between our \texttt{STVA} and other ablation models.}
\label{tab:in-sample-mae}
\resizebox{\linewidth}{!}{%
    \begin{threeparttable}
        \centering
        \begin{tabular}{l:ll:ll:ll:ll}
            \toprule[1pt]\midrule[0.3pt]
            \multicolumn{1}{c:}{} & \multicolumn{1}{c}{MAE} & \multicolumn{1}{c:}{Pct. MAE Increase$^1$} & \multicolumn{1}{c}{\begin{tabular}[c]{@{}c@{}}MAE for \\ \emph{COVID-19 1st Peaks}$^2$\end{tabular}} & \multicolumn{1}{c:}{Pct. MAE Increase} & \multicolumn{1}{c}{\begin{tabular}[c]{@{}c@{}}MAE for\\\emph{COVID-19 2nd Peaks}$^3$\end{tabular}} & \multicolumn{1}{c:}{Pct. MAE Increase} & \multicolumn{1}{c}{\begin{tabular}[c]{@{}c@{}}MAE for\\\emph{Most Affected Region}$^4$\end{tabular}} & \multicolumn{1}{c}{Pct. MAE Increase}\\ \hline
            \texttt{STVA} & \multicolumn{1}{c}{1.63} & \multicolumn{1}{c:}{N/A} & \multicolumn{1}{c}{0.31} & \multicolumn{1}{c:}{N/A} & \multicolumn{1}{c}{2.96} & \multicolumn{1}{c:}{N/A} & \multicolumn{1}{c}{10.35} & \multicolumn{1}{c}{N/A} \\
            \texttt{STVA-spatial} & \multicolumn{1}{c}{1.99} & \multicolumn{1}{c:}{18.09\%} & \multicolumn{1}{c}{0.94} & \multicolumn{1}{c:}{67.02\%} & \multicolumn{1}{c}{3.24} & \multicolumn{1}{c:}{8.64\%} & \multicolumn{1}{c}{14.19} & \multicolumn{1}{c}{27.06\%}\\
            \texttt{STVA-census} & \multicolumn{1}{c}{2.45} & \multicolumn{1}{c:}{33.47\%} & \multicolumn{1}{c}{0.66} & \multicolumn{1}{c:}{53.03\%} & \multicolumn{1}{c}{4.43} & \multicolumn{1}{c:}{33.18\%} & \multicolumn{1}{c}{15.65} & \multicolumn{1}{c}{33.87\%}\\
            \texttt{STVA-mobility} & \multicolumn{1}{c}{5.73} & \multicolumn{1}{c:}{71.55\%} & \multicolumn{1}{c}{2.10} & \multicolumn{1}{c:}{85.23\%} & \multicolumn{1}{c}{7.97} & \multicolumn{1}{c:}{62.86\%} & \multicolumn{1}{c}{12.91} & \multicolumn{1}{c}{19.83\%}\\ 
            \midrule[0.3pt]\bottomrule[1pt]
        \end{tabular}%
        \begin{tablenotes}
            \small
            \item $^1$ Pct. MAE Increase is the percentage of increased MAE resulted by removing one of the components in \texttt{STVA}.
            \item $^2$ \emph{COVID-19 1st Peaks} refers to a certain time period (four weeks from April 5th, 2020 to May 3rd, 2020) that the number of deaths in the U.S. reaches its first peak. 
            \item $^3$ \emph{COVID-19 2nd Peaks} refers to a certain time period (four weeks from December 13th, 2020 to January 3rd, 2021) that the number of deaths in the U.S. reaches its second peak. 
            \item $^4$ \emph{Most Affected Region} refers to the region of eight states with the highest number of deaths due to COVID-19, including New York, New Jersey, California, Texas, Florida, Massachusetts, Illinois, Pennsylvania.
        \end{tablenotes}
    \end{threeparttable}
}
\end{table}

\subsubsection{In-sample estimation}

To evaluate the effectiveness of our method, we compare the county-level \emph{in-sample estimation} on the number of deaths.
The in-sample estimation is a process of evaluating the model's explanatory capabilities using observed data to see how effective the model is in reproducing the data. The process can be carried out as follows:
We first fit the model using the entire data set from March 15th, 2020 to January 17th, 2021, which contains 3,144 counties and 49 weeks in total. 
The in-sample estimation can then be obtained by feeding the same data into the fitted model and recovering the estimation for deaths according to \eqref{eq:case-death-model}.
We also carry out three ablation studies to further investigate the effectiveness of each model component. 
To be specific, we consider three variant models: (i) \texttt{STVA$-$mobility} removes the component of mobility (the second term in \eqref{eq:case-death-model}); (ii) \texttt{STVA$-$census} removes the component of demographic census (the third term in \eqref{eq:case-death-model}); and (iii) \texttt{STVA$-$spatial} simply removes the spatial correlation (the first term in \eqref{eq:case-death-model}) by a diagonal matrix. 
We note that each county can also be regarded as an independent auto-regressive (AR) model without spatial correlation.
We report the in-sample estimation of the proposed approach and the ablation models by aggregating the county-level estimated numbers in the same state. 
As shown in Fig.~\ref{fig:statewise-death}, we select eight major states with the highest total number of deaths in the U.S. The shaded area indicates the actual number of deaths reported in the COVID-19 data set, and the solid red line indicates the in-sample estimated deaths by our model. 
We observe that the \texttt{STVA} (solid red lines) can capture the dynamics of true death trajectories better than the other three ablation models (dash lines). 
We also provide quantitative results for county-level MAE in Table~\ref{tab:in-sample-mae}. 
As we can see, the result indicates that our model \texttt{STVA} generally attains the lowest MAE for all scenarios. 
There are also significant performance gains if we only focus on predicting the peak week and the most affected region in the U.S.
The result suggests that these components are conducive to improving the model's performance.
In Appendix~\ref{append:other-results}, we also present the in-sample estimation of the confirmed cases for the same eight states.

\begin{figure}[!t]
\centering
\begin{subfigure}[h]{.325\linewidth}
\includegraphics[width=\textwidth]{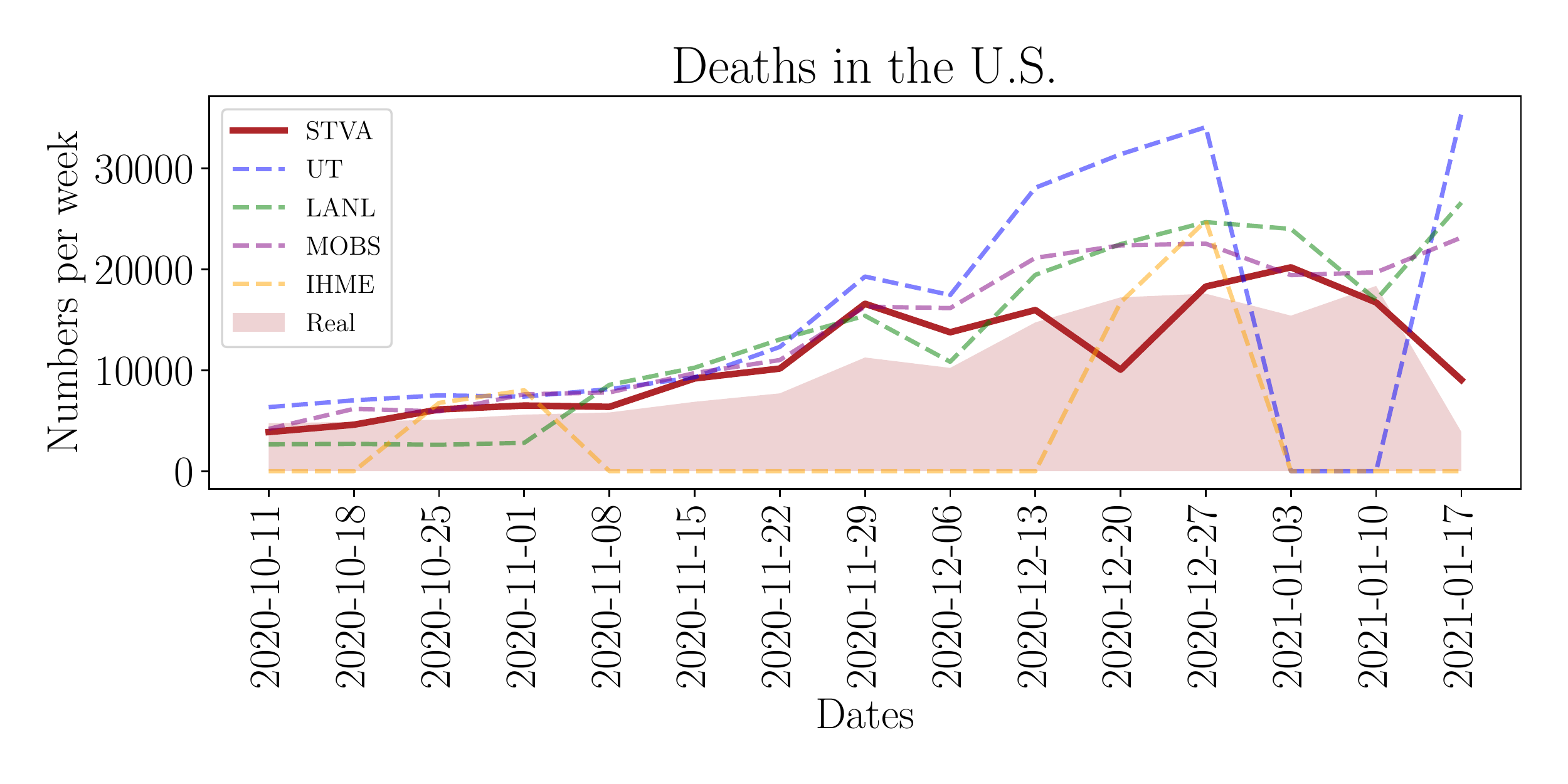}
\end{subfigure}
\begin{subfigure}[h]{.325\linewidth}
\includegraphics[width=\textwidth]{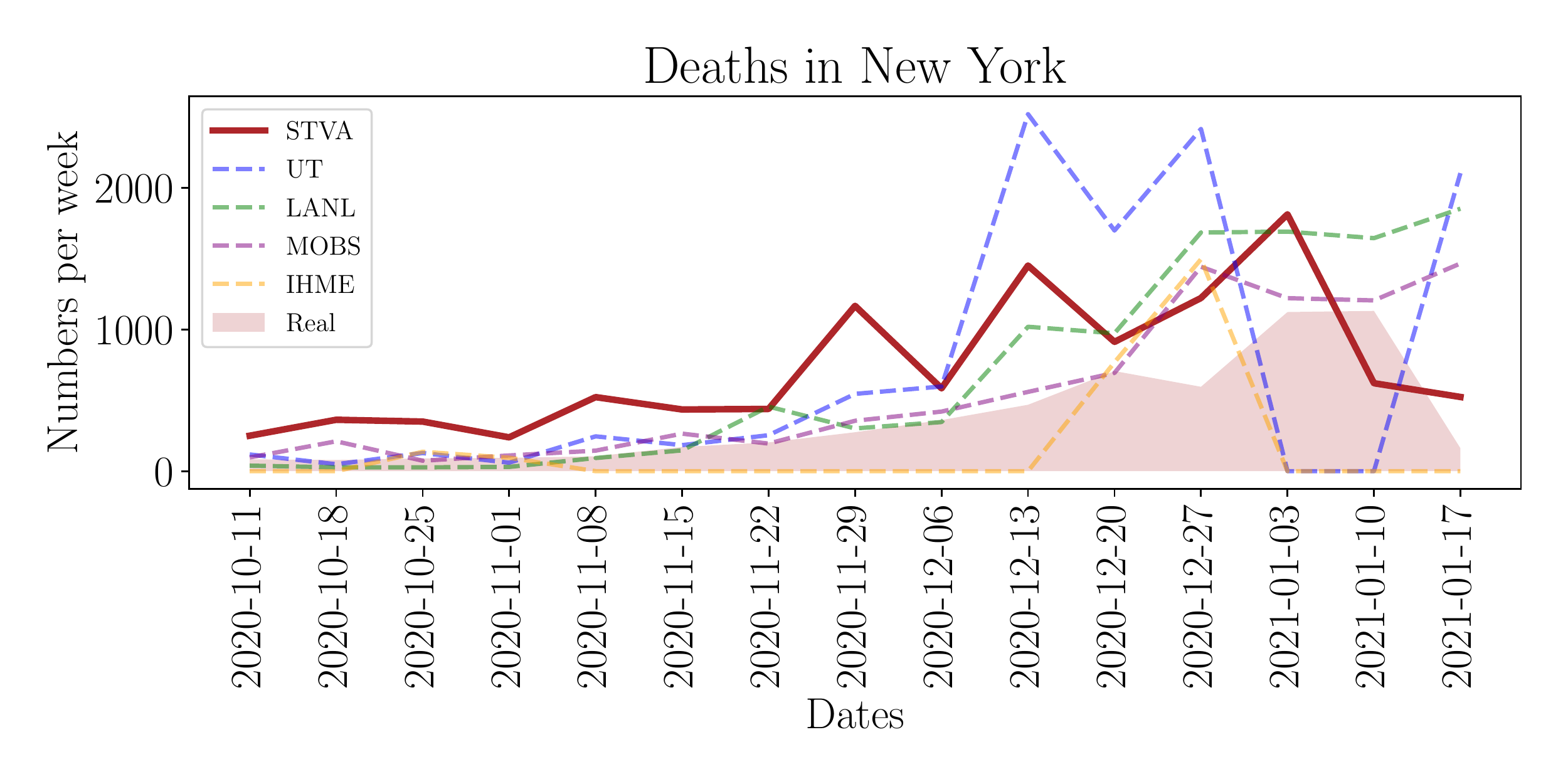}
\end{subfigure}
\begin{subfigure}[h]{.325\linewidth}
\includegraphics[width=\textwidth]{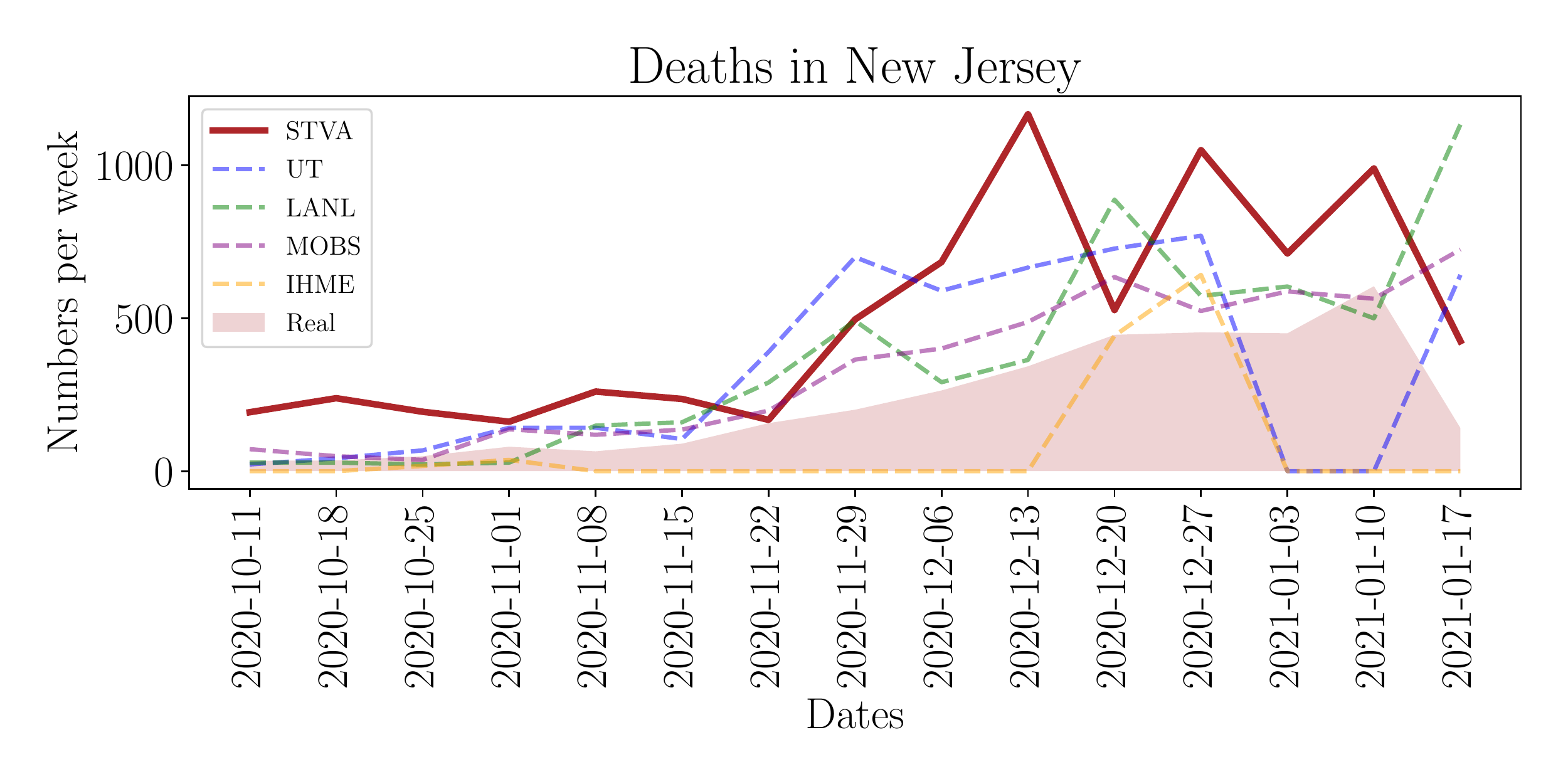}
\end{subfigure}
\vfill
\begin{subfigure}[h]{.325\linewidth}
\includegraphics[width=\textwidth]{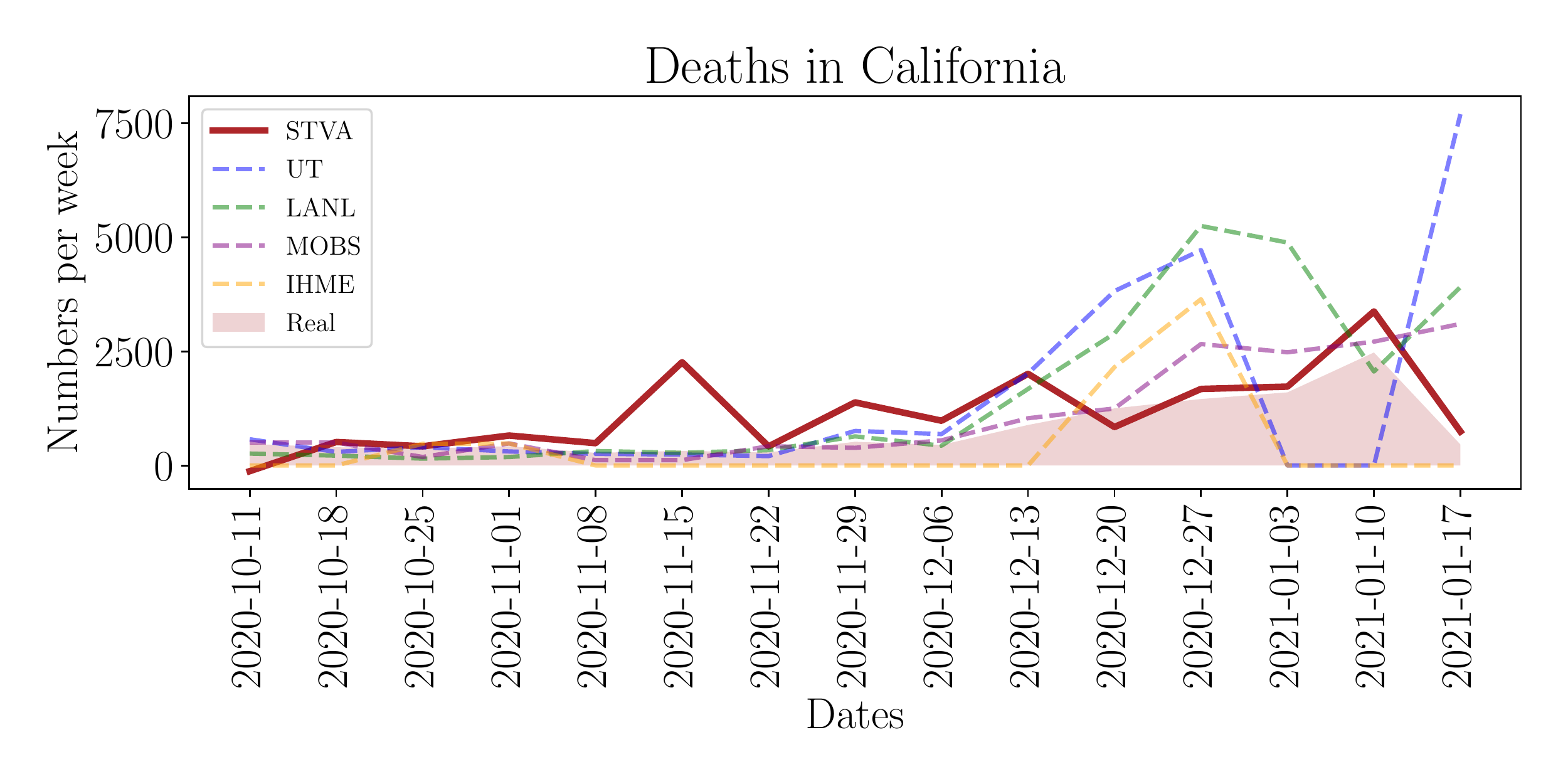}
\end{subfigure}
\begin{subfigure}[h]{.325\linewidth}
\includegraphics[width=\textwidth]{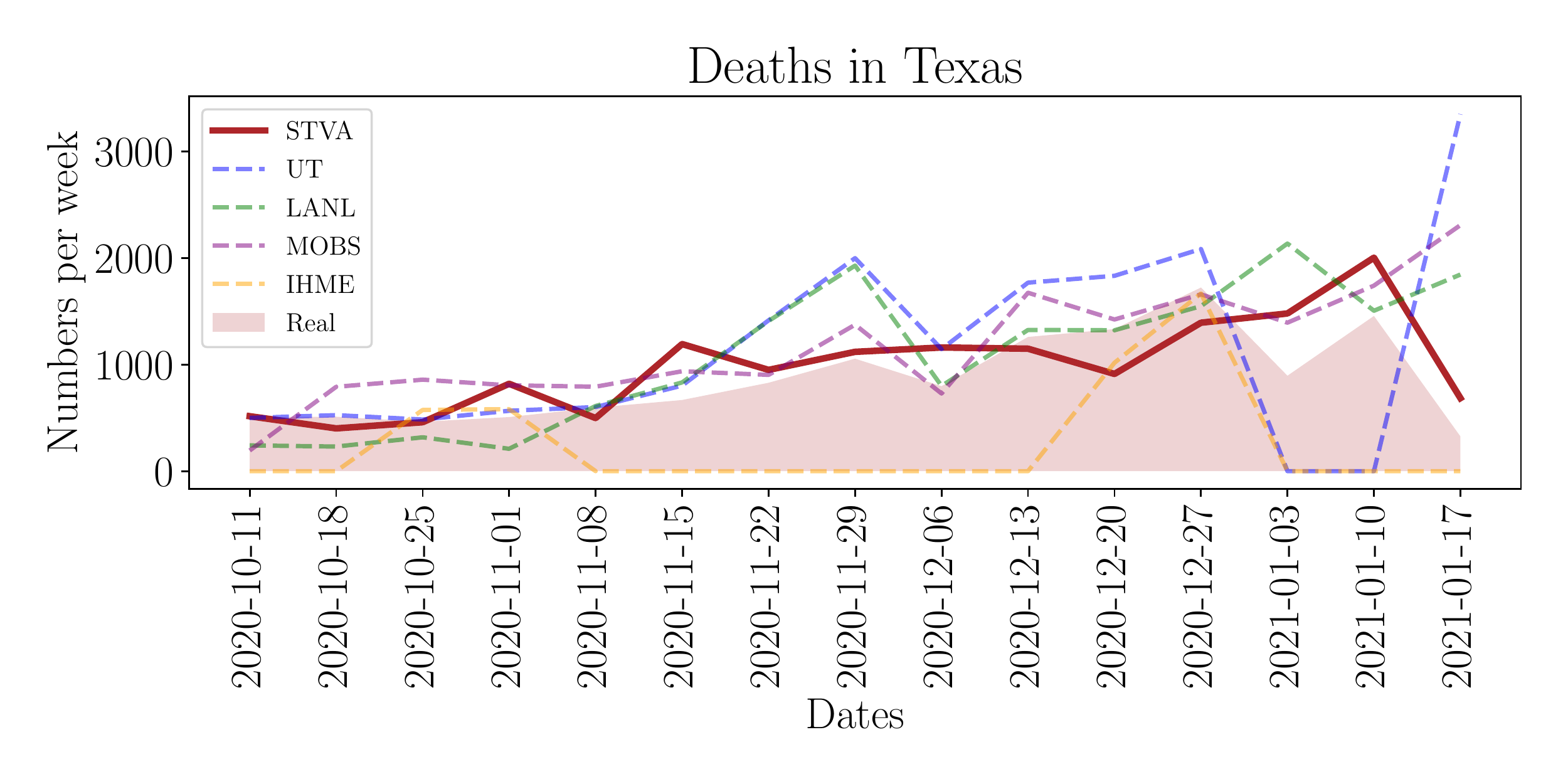}
\end{subfigure}
\begin{subfigure}[h]{.325\linewidth}
\includegraphics[width=\textwidth]{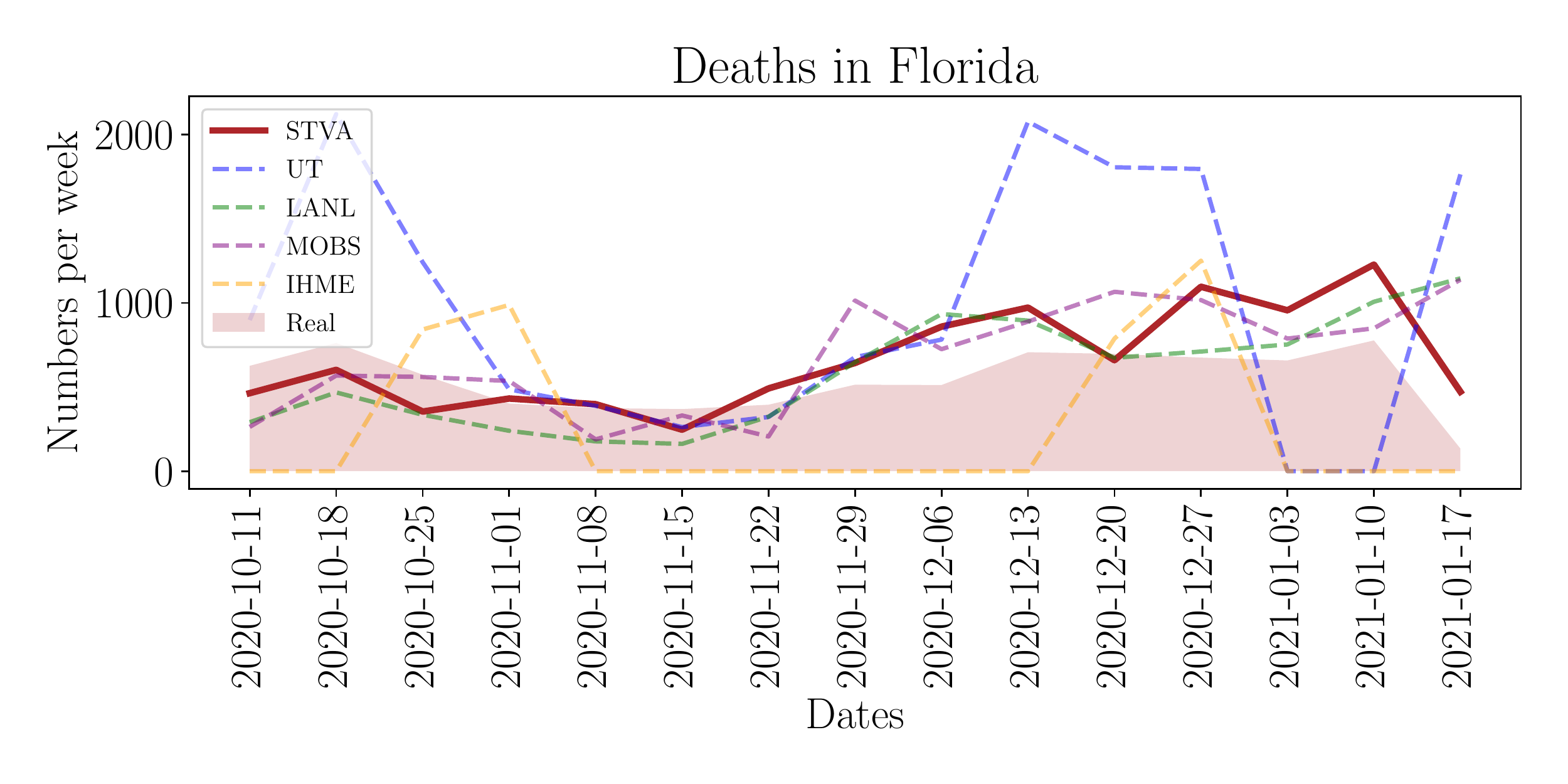}
\end{subfigure}
\vfill
\begin{subfigure}[h]{.325\linewidth}
\includegraphics[width=\textwidth]{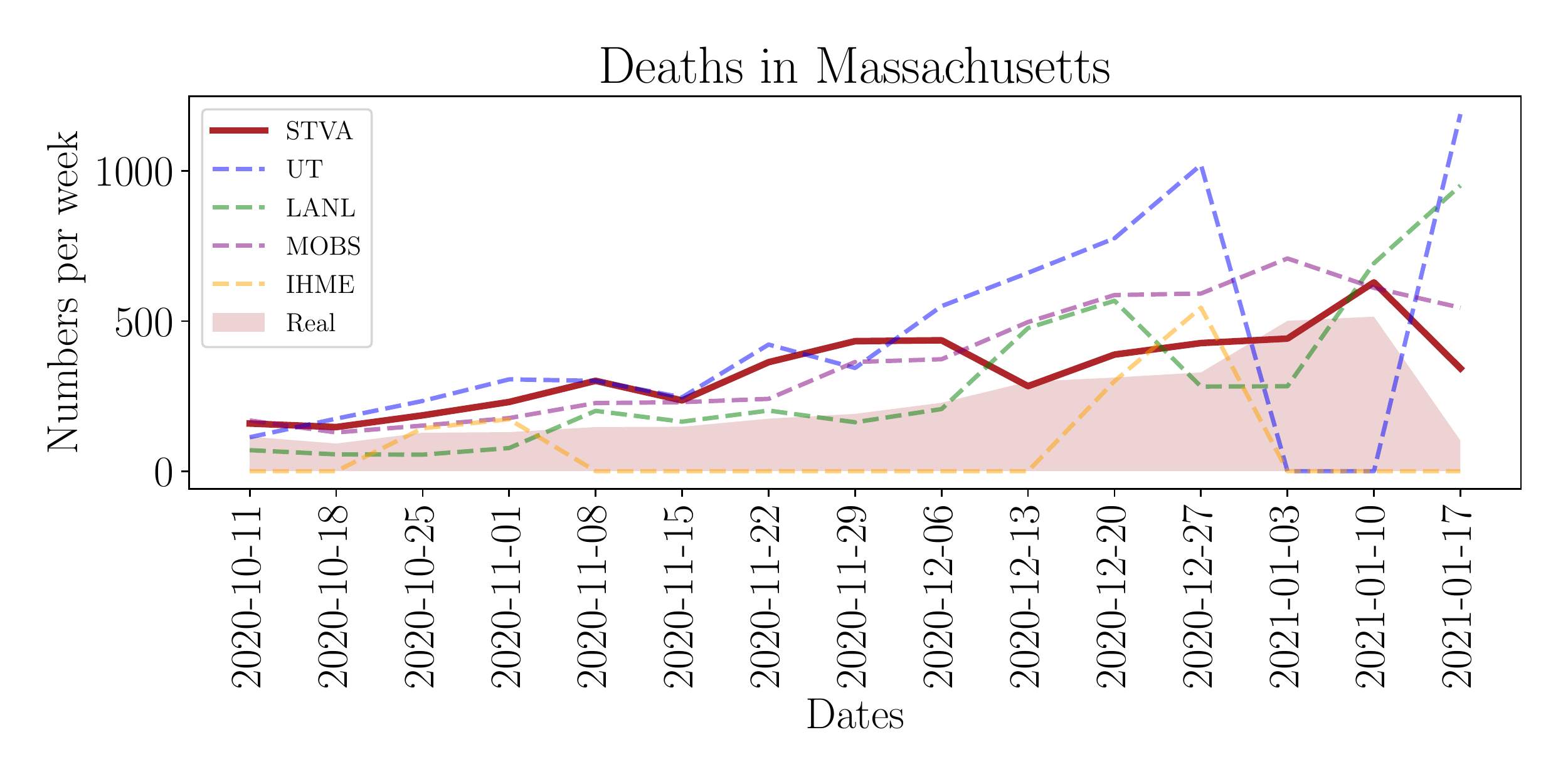}
\end{subfigure}
\begin{subfigure}[h]{.325\linewidth}
\includegraphics[width=\textwidth]{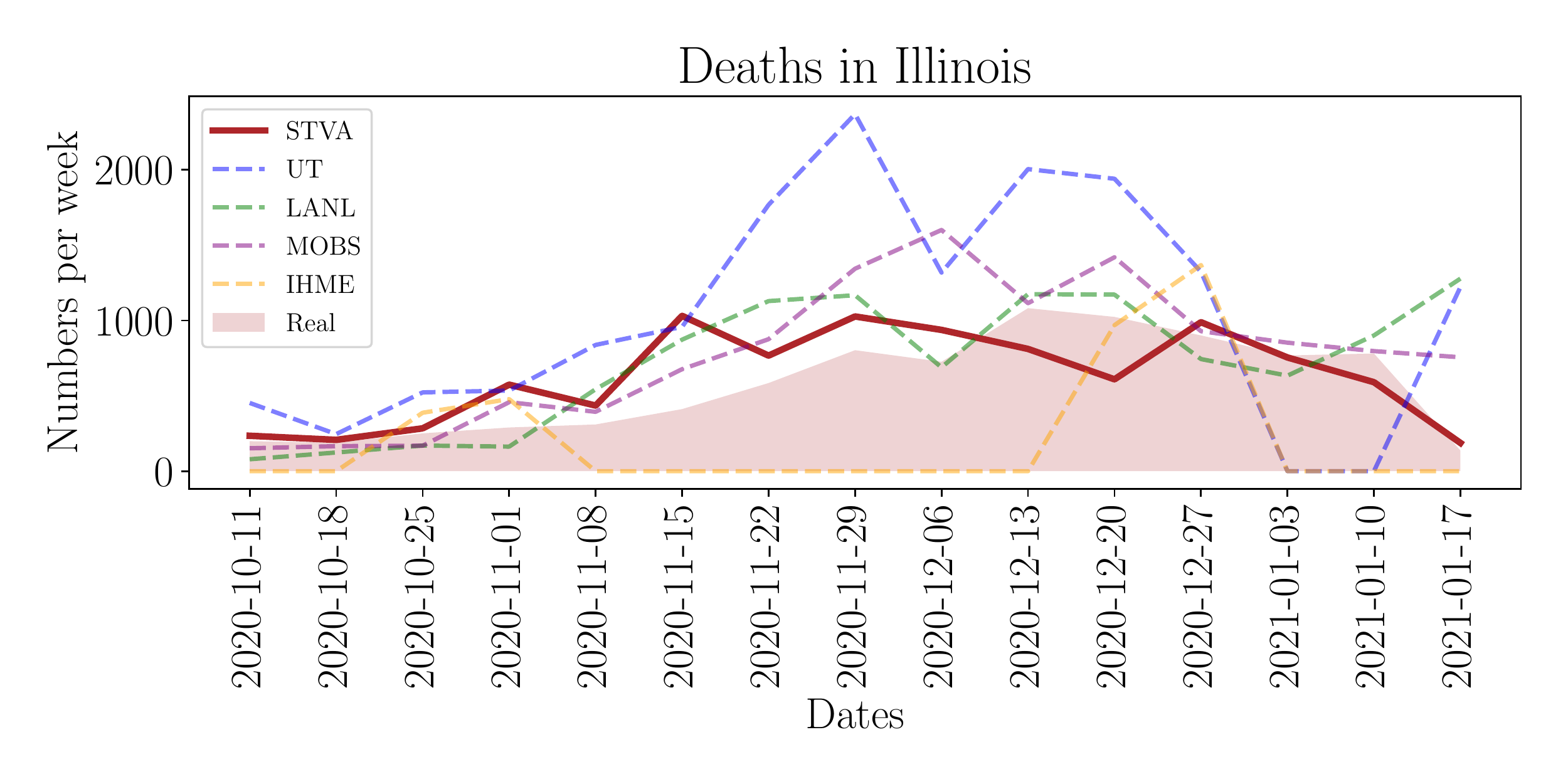}
\end{subfigure}
\begin{subfigure}[h]{.325\linewidth}
\includegraphics[width=\textwidth]{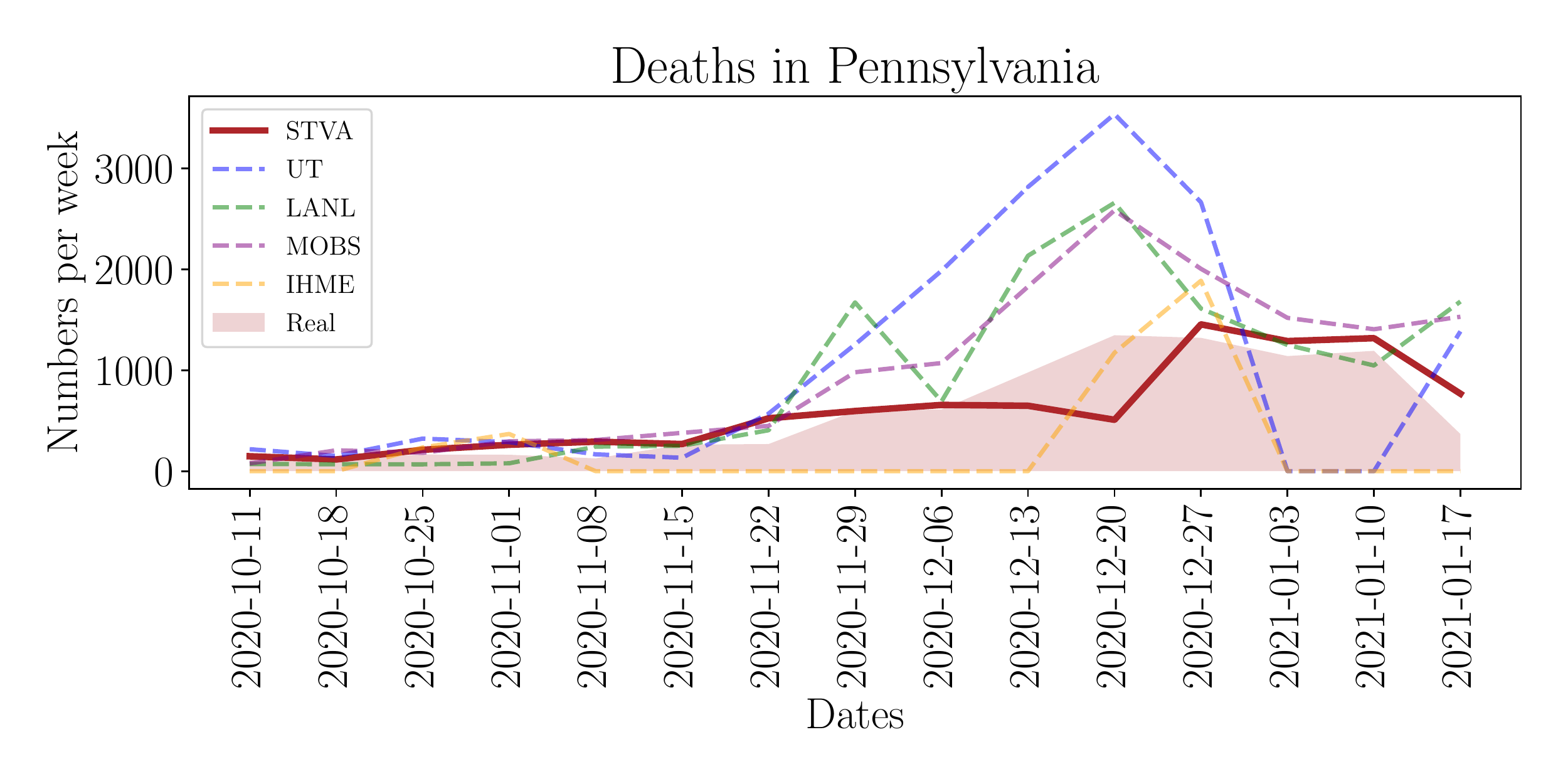}
\end{subfigure}
\vspace{-.1in}
\caption{One-week ahead death prediction for the U.S. and other eight major states with the highest number of COVID-19 deaths in the U.S. The \texttt{STVA}'s statewise predictions are obtained by aggregating the county-level predictions in the same states. Figures are sorted in descending order of the total number of confirmed cases since October 11th, 2020. The results show that our model can achieve promising predictive performance with other four benchmarks. Note that these four benchmarks only provide state-level predictions.}
\label{fig:outofsample-statewise-death}
\vspace{-.1in}
\end{figure}

\subsubsection{Out-of-sample prediction}

In addition to the in-sample estimation, we assess the model's predictive power by performing the \emph{one-week ahead (out-of-sample) prediction}.
The prediction procedure withholds the future data from the model estimation, then uses the fitted model to make predictions for the (hold-out) data in the next week.

Here, we compare our model against four benchmark methods adopted by CDC, which represent the current state-of-art for COVID-19 prediction: (1) COVID-19 Mortality Projections for the U.S. States by the University of Texas, Austin (\texttt{UT}) \cite{UTaustin, woody2020projections}: they introduce a negative-binomial mixed-effects generalized linear model (GLM), i.e., the predictor is a GLM with a logarithm link function.; (2) COVID-19 Cases and Deaths Forecasts by the Los Alamos National Laboratory (\texttt{LANL}) \cite{LANL}: the model assumes that a fraction of the newly generated cases will die and proposes a statistical model to capture this effect; (3) COVID-19 Modeling by the Northeastern University, Laboratory for the Modeling of Biological and Socio-technical Systems (\texttt{MOBS}) \cite{MOBS}: their team adopts a classic SLIR (Susceptible-Latent-Infectious-Removed)-like compartmentalization scheme for disease progression; (4) COVID-19 Projections for the United States by the Institute for Health Metrics and Evaluation (\texttt{IHME}) \cite{IHME, covid2021modeling}: this project considers a deterministic SEIR compartmental framework.
In particular, all these four approaches mainly use the state-level records of the number of cases and deaths as the input of their models. At the same time, \texttt{IHME} also takes advantage of several critical driving covariates (pneumonia seasonality, mobility, testing rates, and mask use per capita).
The prediction results of these benchmark methods are directly quoted from the CDC's official reports \cite{CDC}.

We only present the results from October 11th, 2020, for one-week ahead death prediction. The statistics before October are inaccurate due to the low testing rate, and the data are insufficient to fit the model. 
Similar to the in-sample estimation, we report the prediction results for the entire U.S. and eight top states with the highest number of deaths in Fig.~\ref{fig:outofsample-statewise-death}.
It has shown that the aggregated county-level predictions suggested by the \texttt{STVA} (solid green lines) achieve competitive performance against other mainstream approaches (dash lines).
It is worth noting that these four methods only provide state-level forecasting for the number of deaths, which is less challenging than the county-level prediction.

\begin{figure}[!t]
\centering
\begin{subfigure}[h]{.47\linewidth}
\includegraphics[width=\textwidth]{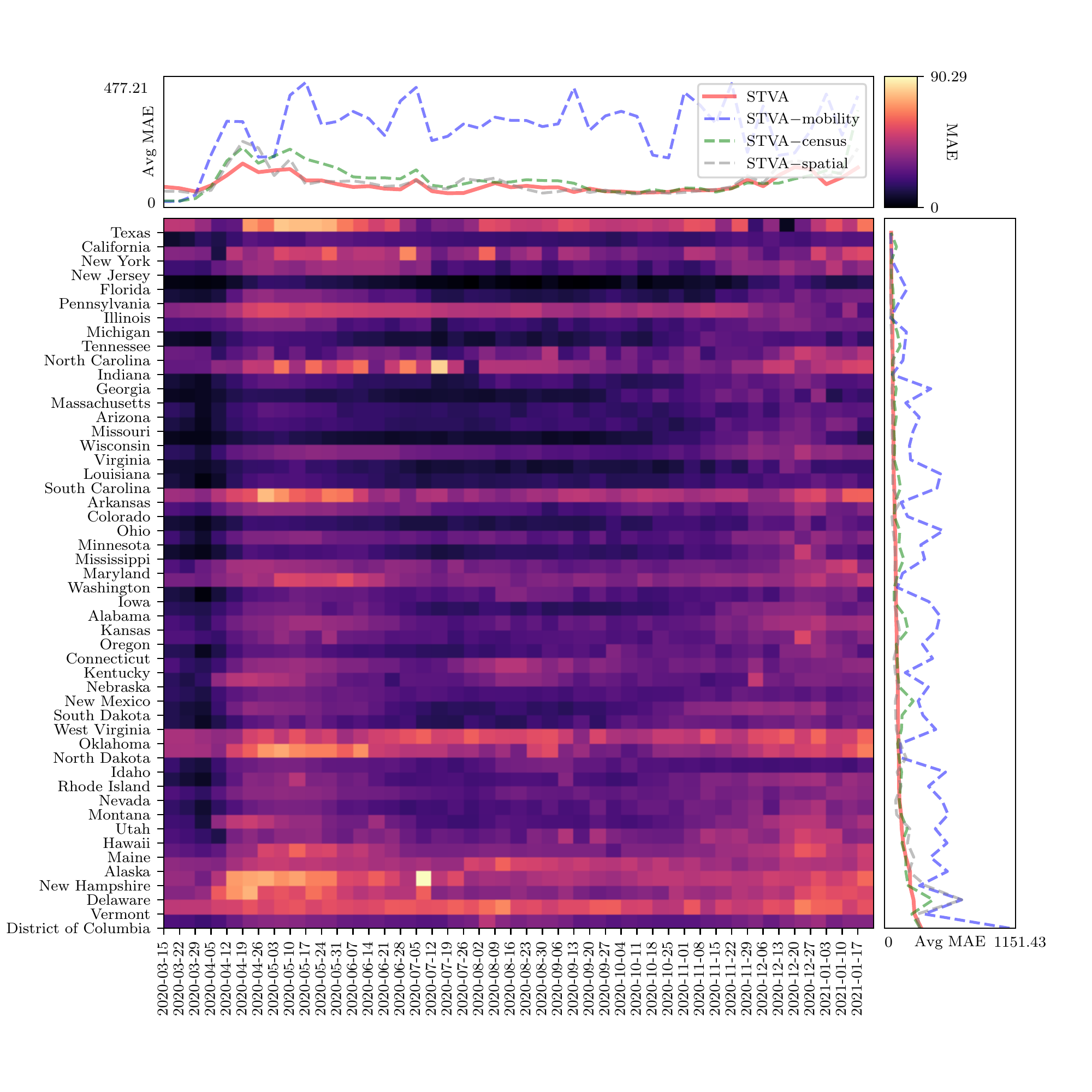}
\caption{in-sample estimation}
\end{subfigure}
\begin{subfigure}[h]{.47\linewidth}
\includegraphics[width=\textwidth]{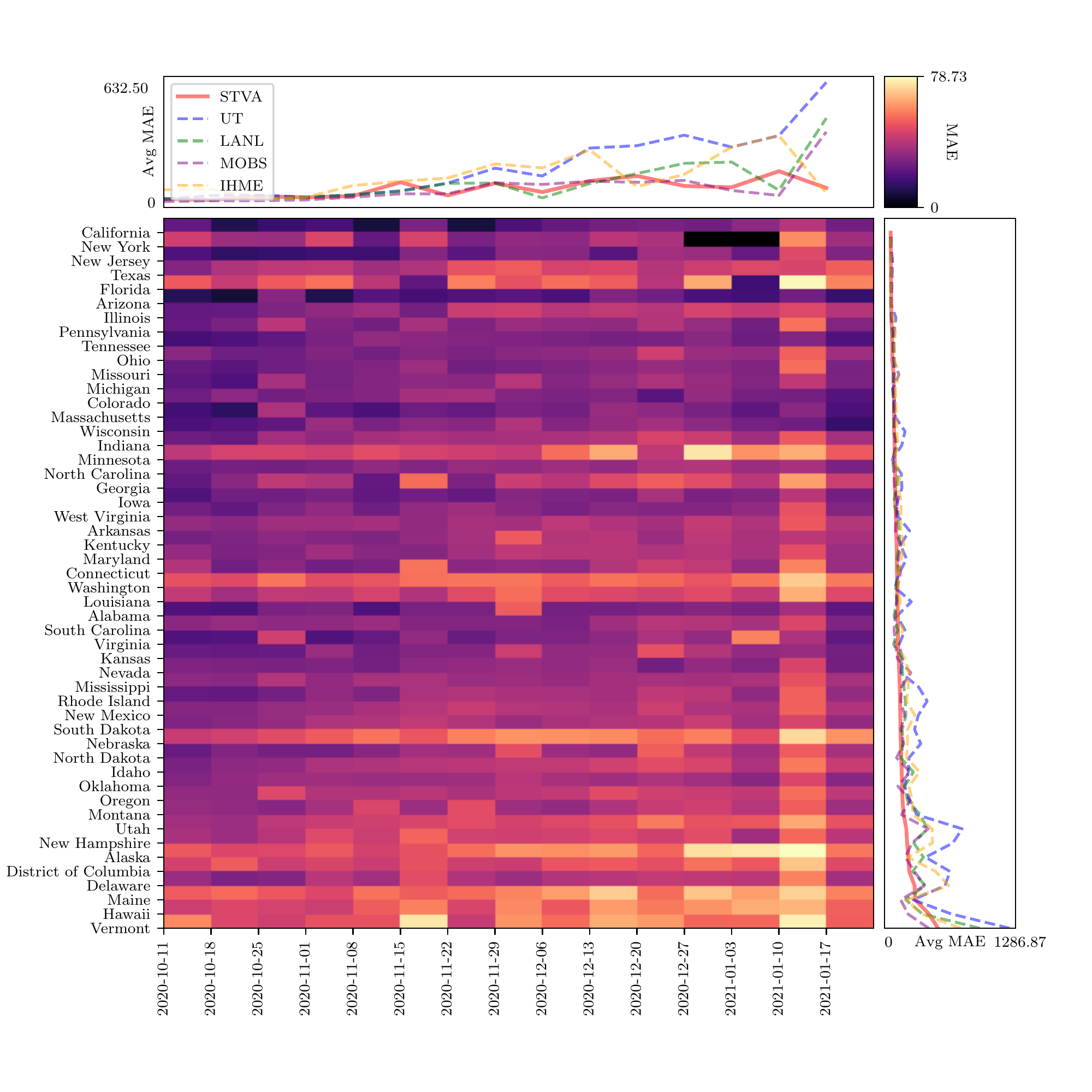}
\caption{one-week ahead (out-of-sample) prediction}
\end{subfigure}
\caption{Mean absolute error (MAE) of (a) the in-sample death estimation and (b) the one-week ahead out-of-sample death prediction by our model. The color depth of the heatmaps indicates the MAE of county-level prediction for certain state and week. The horizontal and vertical line charts show the average MAE over weeks and states comparing to other benchmark methods. The states have been sorted in the ascending order of their MAE from top to bottom. Note that four benchmarks in (b) only provide state-level predictions; We compare our method against these benchmarks by summing up the deaths numbers in the same state.}
\label{fig:mae}
\vspace{-.1in}
\end{figure}

More quantitative results are summarized in Fig.~\ref{fig:mae}. The heatmaps show the mean absolute error (MAE) of the county-level estimation/prediction within a particular state and at certain weeks. The average MAEs over states and weeks are presented in the vertical line chart on the right and the horizontal line chart on the top. The states are sorted in the ascending order of their MAE from top to bottom. 
As shown in Fig.~\ref{fig:mae}, our model significantly outperforms the other ablation models while achieving competitive predictive performances compared to the other widely-adopted state-level approaches.
We can also observe that: our model tends to achieve better performance for the states with larger populations, such as Florida, New York, Texas, et cetera; for the deaths, our model has a balanced performance in each state, and the MAE is getting better (smaller) and becoming more stable after the summer surge of the COVID-19 (from June to July). 

\subsection{Model interpretation}

Our study focuses on exploring the in-sample explanatory content of predetermined factors in our model. We fit the model using the entire data set collected from three data mentioned above sources in Section~\ref{sec:data}, and interpret the model by examining its fitted coefficients. 

\begin{figure}[h]
\centering
\begin{subfigure}[h]{.25\linewidth}
\includegraphics[width=\textwidth]{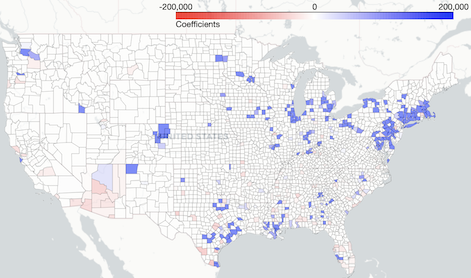}
\caption{Chicago in $\boldsymbol A_1$}
\end{subfigure}
\begin{subfigure}[h]{.25\linewidth}
\includegraphics[width=\textwidth]{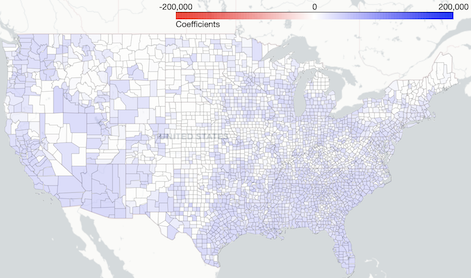}
\caption{Chicago in $\boldsymbol B_1$}
\end{subfigure}
\begin{subfigure}[h]{.25\linewidth}
\includegraphics[width=\textwidth]{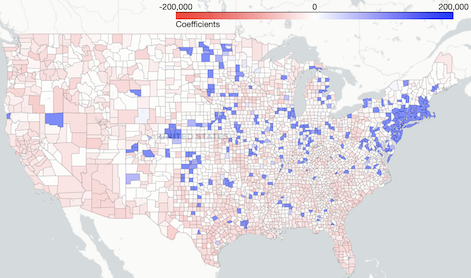}
\caption{Chicago in $\boldsymbol H_1$}
\end{subfigure}
\vfill
\begin{subfigure}[h]{.25\linewidth}
\includegraphics[width=\textwidth]{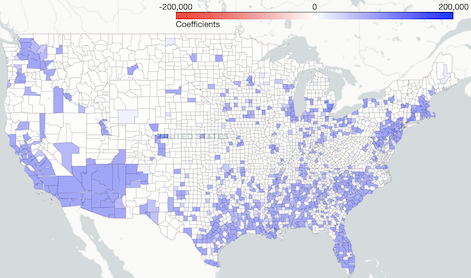}
\caption{New York in $\boldsymbol A_1$}
\end{subfigure}
\begin{subfigure}[h]{.25\linewidth}
\includegraphics[width=\textwidth]{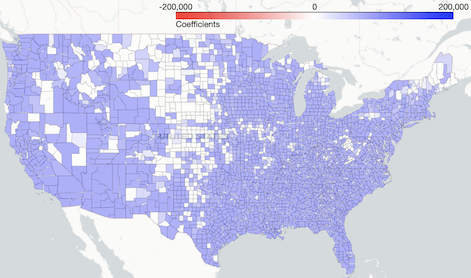}
\caption{New York in $\boldsymbol B_1$}
\end{subfigure}
\begin{subfigure}[h]{.25\linewidth}
\includegraphics[width=\textwidth]{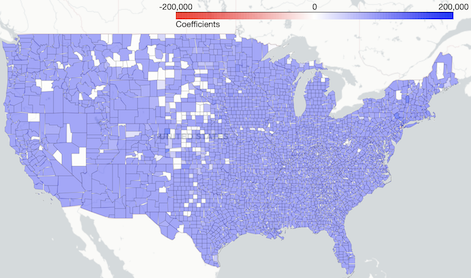}
\caption{New York in $\boldsymbol H_1$}
\end{subfigure}
\vfill
\begin{subfigure}[h]{.25\linewidth}
\includegraphics[width=\textwidth]{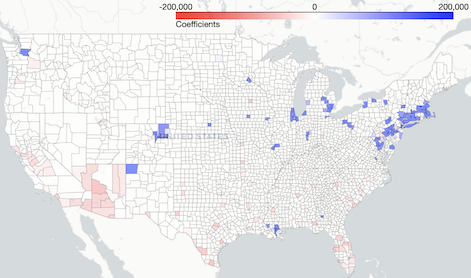}
\caption{Dallas in $\boldsymbol A_1$}
\end{subfigure}
\begin{subfigure}[h]{.25\linewidth}
\includegraphics[width=\textwidth]{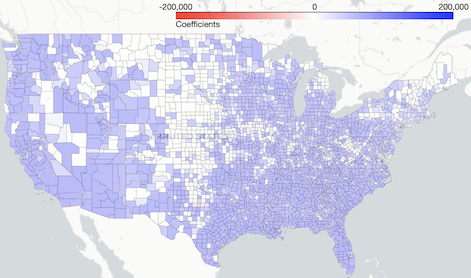}
\caption{Dallas in $\boldsymbol B_1$}
\end{subfigure}
\begin{subfigure}[h]{.25\linewidth}
\includegraphics[width=\textwidth]{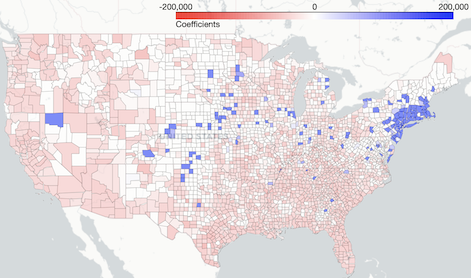}
\caption{Dallas in $\boldsymbol H_1$}
\end{subfigure}
\vfill
\begin{subfigure}[h]{.25\linewidth}
\includegraphics[width=\textwidth]{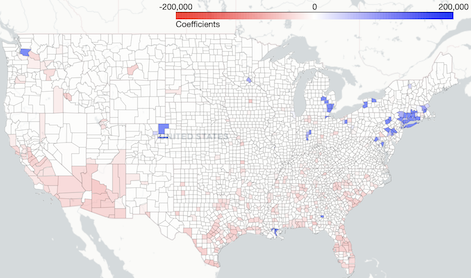}
\caption{Houston in $\boldsymbol A_1$}
\end{subfigure}
\begin{subfigure}[h]{.25\linewidth}
\includegraphics[width=\textwidth]{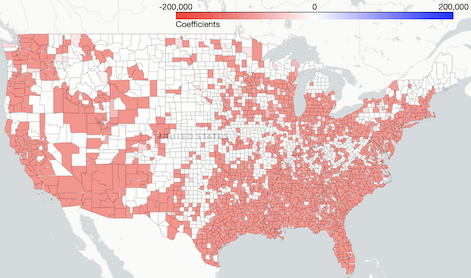}
\caption{Houston in $\boldsymbol B_1$}
\end{subfigure}
\begin{subfigure}[h]{.25\linewidth}
\includegraphics[width=\textwidth]{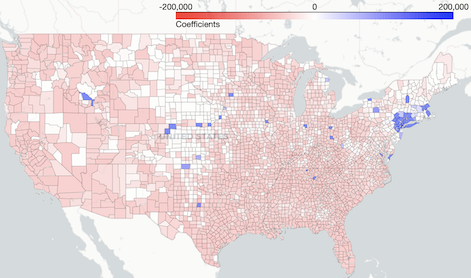}
\caption{Houston in $\boldsymbol H_1$}
\end{subfigure}
\vfill
\begin{subfigure}[h]{.25\linewidth}
\includegraphics[width=\textwidth]{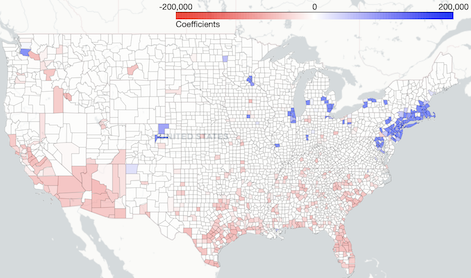}
\caption{Miami in $\boldsymbol A_1$}
\end{subfigure}
\begin{subfigure}[h]{.25\linewidth}
\includegraphics[width=\textwidth]{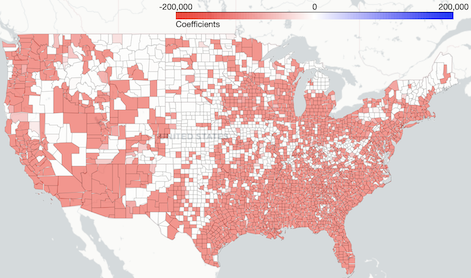}
\caption{Miami in $\boldsymbol B_1$}
\end{subfigure}
\begin{subfigure}[h]{.25\linewidth}
\includegraphics[width=\textwidth]{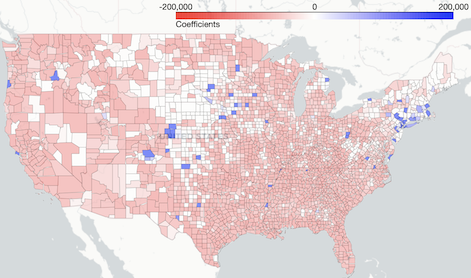}
\caption{Miami in $\boldsymbol H_1$}
\end{subfigure}
\caption{Examples of coefficients for hubs in matrix $\boldsymbol \Lambda_1$. 
For instance, the color depth of any county $i$ in (a) represents the value of coefficient $\alpha_{i,j,1}$ in $\boldsymbol A_1$, where county $j$ is Chicago. Counties in blue indicate their current number of deaths is positively related to its number of deaths in the last week; counties in red are the opposite; counties in white represent no discernable correlation between the two numbers.
Coefficients of different hubs show the various spatial pattern in ``spreading'' or ``controlling'' the disease. 
}
\label{fig:spatial-coeff}
\vspace{-.2in}
\end{figure}

\begin{figure}[t]
\centering
\begin{subfigure}[h]{.25\linewidth}
\includegraphics[width=\textwidth]{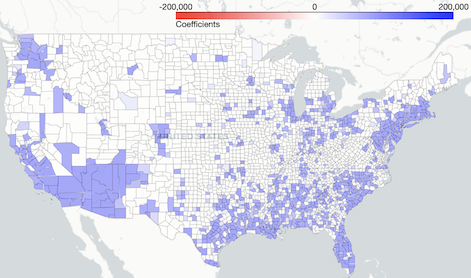}
\caption{Atlanta in $\boldsymbol A_1$}
\end{subfigure}
\begin{subfigure}[h]{.25\linewidth}
\includegraphics[width=\textwidth]{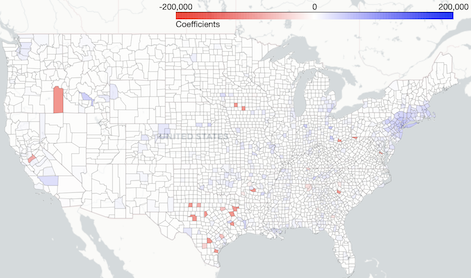}
\caption{Seattle in $\boldsymbol B_1$}
\end{subfigure}
\begin{subfigure}[h]{.25\linewidth}
\includegraphics[width=\textwidth]{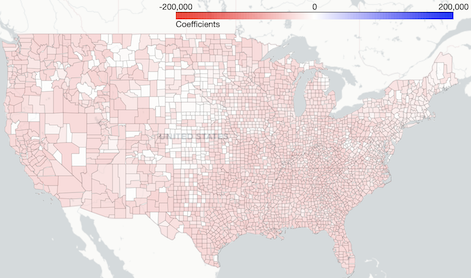}
\caption{Los Angeles in $\boldsymbol H_1$}
\end{subfigure}
\vfill
\begin{subfigure}[h]{.25\linewidth}
\includegraphics[width=\textwidth]{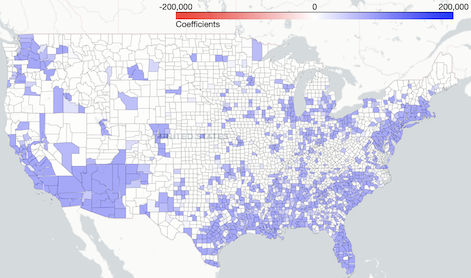}
\caption{Atlanta in $\boldsymbol A_2$}
\end{subfigure}
\begin{subfigure}[h]{.25\linewidth}
\includegraphics[width=\textwidth]{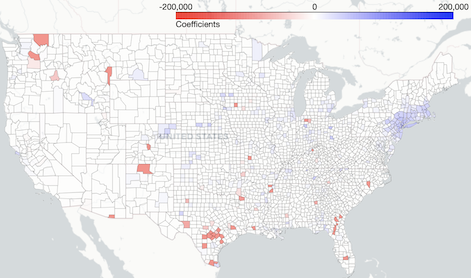}
\caption{Seattle in $\boldsymbol B_2$}
\end{subfigure}
\begin{subfigure}[h]{.25\linewidth}
\includegraphics[width=\textwidth]{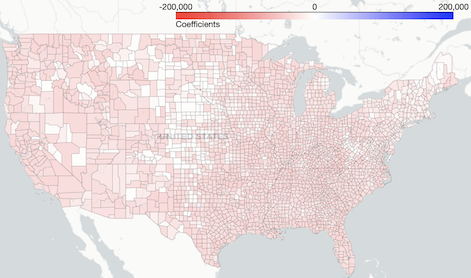}
\caption{Los Angeles in $\boldsymbol H_2$}
\end{subfigure}
\caption{Examples of coefficients with different time lag. }
\label{fig:temporal-coeff}
\vspace{-.1in}
\end{figure}

\begin{figure}[t]
\centering
\begin{subfigure}[h]{.25\linewidth}
\includegraphics[width=\textwidth]{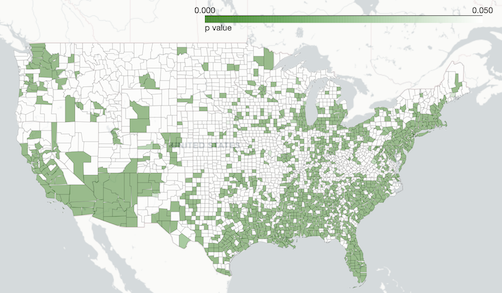}
\caption{Atlanta}
\end{subfigure}
\begin{subfigure}[h]{.25\linewidth}
\includegraphics[width=\textwidth]{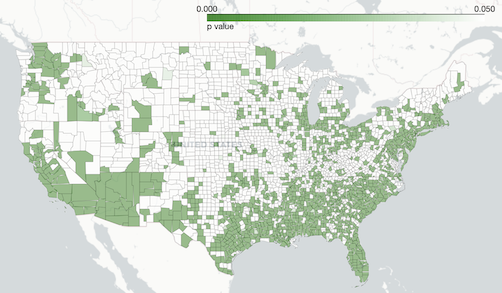}
\caption{Los Angeles}
\end{subfigure}
\begin{subfigure}[h]{.25\linewidth}
\includegraphics[width=\textwidth]{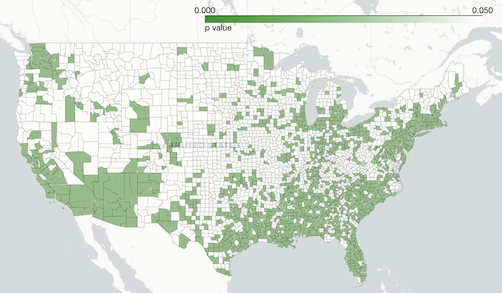}
\caption{Washington D.C.}
\end{subfigure}
\vfill
\begin{subfigure}[h]{.25\linewidth}
\includegraphics[width=\textwidth]{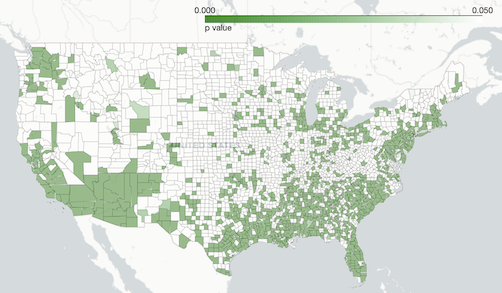}
\caption{New York}
\end{subfigure}
\begin{subfigure}[h]{.25\linewidth}
\includegraphics[width=\textwidth]{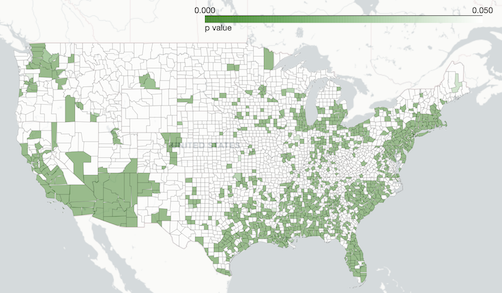}
\caption{Seattle}
\end{subfigure}
\begin{subfigure}[h]{.25\linewidth}
\includegraphics[width=\textwidth]{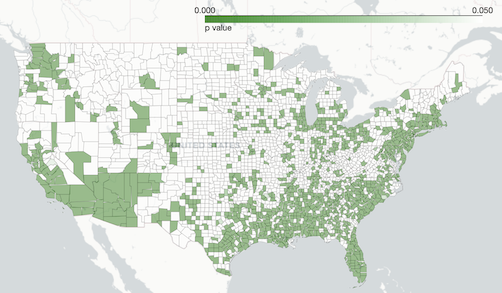}
\caption{San Francisco}
\end{subfigure}
\vfill
\begin{subfigure}[h]{.25\linewidth}
\includegraphics[width=\textwidth]{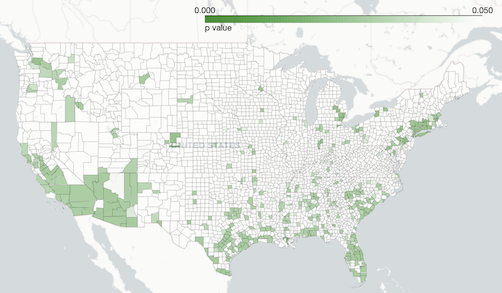}
\caption{Houston}
\end{subfigure}
\begin{subfigure}[h]{.25\linewidth}
\includegraphics[width=\textwidth]{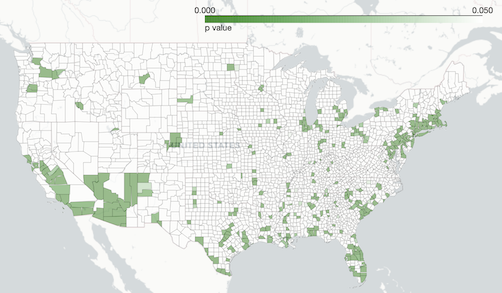}
\caption{Dallas}
\end{subfigure}
\begin{subfigure}[h]{.25\linewidth}
\includegraphics[width=\textwidth]{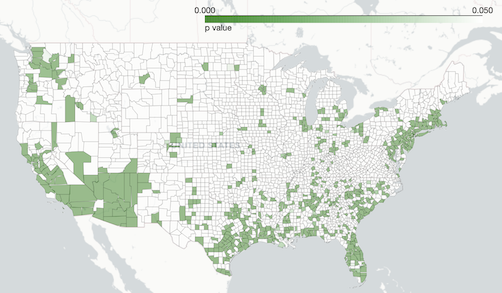}
\caption{Miami}
\end{subfigure}
\caption{Examples of $p$-values for hubs' coefficients in $\boldsymbol{A}_1$.}
\label{fig:p-value-A0}
\vspace{-.1in}
\end{figure}


\paragraph{Spatio-temporal dependencies between cases and deaths}

The experimental results demonstrate a distinctive underlying spatio-temporal pattern between confirmed cases and deaths of the COVID-19.
We first report the coefficients of five representative hubs in $\boldsymbol \Lambda_1$ in Fig.~\ref{fig:spatial-coeff}. 
The hubs' coefficients in $\boldsymbol B_1$, $\boldsymbol H_1$, $\boldsymbol A_1$ reveal their spatial dependencies between each pair of past cases and current cases, past cases and current deaths, and past deaths and current deaths, respectively. 
As we can see, hubs have a strong ``radiating'' power on most of the United States regions and contribute a great deal to promote or curb the spread of the COVID-19.
However, the rural area with lower population density in the Central United States is not significantly influenced by the hubs.  
The hubs situated in the Northern United States (e.g., Chicago, New York) are negatively related to the spreading diseases to the other regions (in blue), which appear to have better controls on the expansion of the virus. 
In contrast, the hubs in the Southern United States (e.g., Dallas, Houston, Miami) usually are positively related to the increases of both cases and deaths in other regions (in red).
The result also presents some other interesting findings: some hubs show two opposite influences on the cases and deaths in the same region. For example,  we see that on the one hand, Fig.~\ref{fig:spatial-coeff} (m) shows that the number of deaths in Miami is negatively related to the deaths in the New England area of the United States. On the other hand, Fig.~\ref{fig:spatial-coeff} (n) shows that the number of cases in Miami is positively related to the cases in the same area. 
Some hubs contribute to the increase of cases or deaths in one region, reducing the cases or deaths in other regions. For example, Fig.~\ref{fig:spatial-coeff} (g) and (j) show that Dallas and Houston have a positive impact on the New England area in the United States and have a negative impact on Florida and California. 
Apart from analyzing the spatial structure across regions learned by the model, we also study the temporal dependences in the past two weeks for the same hub.
In Fig.~\ref{fig:temporal-coeff}, we present three typical pairs of comparisons for coefficients between one-week lag and two-week lag: coefficients of Atlanta in $\boldsymbol A_1$ and $\boldsymbol A_2$, coefficients of Seattle in $\boldsymbol B_1$ and $\boldsymbol B_2$, and coefficients of Los Angeles in $\boldsymbol H_1$ and $\boldsymbol H_2$. All three comparisons share one thing in common: coefficients of different time lag have a similar spatial pattern, but the overall coefficients of two-week lag are relatively smaller than corresponding ones of one-week lag. This indicates that the last week has a stronger influence.
We also observe that the spatial pattern in $\boldsymbol{A}$, $\boldsymbol{B}$, and $\boldsymbol{H}$ are significantly different from each other according to Fig.~\ref{fig:spatial-coeff} and Fig.~\ref{fig:temporal-coeff}. This observation confirms that each spatial component described by matrices $\boldsymbol{A}$, $\boldsymbol{B}$, and $\boldsymbol{H}$ plays a different role in capturing the spatial correlation. For example, we can observe that the visualization of the matrix $\boldsymbol{H}$ always shows large positive coefficients in New England area while the coefficients in matrix $\boldsymbol{B}$ in the same area being typically more negligible or even negative. This may be related to the high death rate in the New England area \cite{deathrate} since $\boldsymbol{H}$ captures the spatial correlation between previous cases and future deaths, and $\boldsymbol{B}$ only concerns the similar correlation between cases.
Last but not least,
we further investigate the $p$-values of these spatial coefficients, as finding statistically significant relationships between the prediction and the observation is of great importance to the model evaluation. Since $\boldsymbol A_1$ plays a key role in predicting future deaths by being connected to the death observations in the past, we take the hubs' coefficients in $\boldsymbol A_1$ as an example.
As shown in Fig.~\ref{fig:p-value-A0}, the number of deaths in these hubs is statistically significant to the death counts in other populated regions in the U.S. (New England, the Southeastern United States, and the Western united states). In particular, we find the hubs 
in the west and the north (Los Angeles, Washington D.C., New York, Seattle, and San Francisco) are statistically significant to a broader area than the hubs in the South (Houston, Dallas, and Miami).


\begin{table}[!t]
\centering
\caption{Summary of fitted coefficients for mobility and demographic factors.}
\vspace{-.1in}
\label{tab:covariate-coeff}
\resizebox{\textwidth}{!}{%
\begin{tabular}{lllllllllllllll}
\toprule[1pt]\midrule[0.3pt]
Lag & \multicolumn{6}{l}{One-week} & \multicolumn{6}{l}{Two-week} & N/A & N/A \\ 
Category & Workplaces & Recreation & Grocery & Park & Transit & Residential & Workplaces & Recreation & Grocery & Park & Transit & Residential & Population & Over 65 \\ \hline
Term {\it w.r.t.} case & $\mu_{1,1}$ & $\mu_{2,1}$ & $\mu_{3,1}$ & $\mu_{4,1}$ & $\mu_{5,1}$ & $\mu_{6,1}$ & $\mu_{1,2}$ & $\mu_{2,2}$ & $\mu_{3,2}$ & $\mu_{4,2}$ & $\mu_{5,2}$ & $\mu_{6,2}$ & $\upsilon_{1}$ & $\upsilon_{2}$ \\ \hline
Coefficient & {\bf +9.67e+2} & {\bf +1.67e+3} & -1.21e+3 & -7.83e+2 & {\bf +3.54e+2} & -5.08e+3 & {\bf +1.28+e3} & {\bf +2.16e+3} & -8.52e+2 & -7.25e+2 & {\bf +3.12e+3} & -5.05e+3 & {\bf +2.91e+4} & -1.58e+01 \\
$p$-value & +5.01e-5 & +3.21e-5 & +1.03e-7 & +7.46e-5 & +2.01e-5 & +2.56e-5 & +2.19e-5 & +2.30e-5 & +3.77e-7 & +3.91e-7 & +1.42e-7 & +6.64e-7 & +9.10e-9 & +1.81e-1  \\
$t$-value   & +5.78e+1 & +7.63e+1 & +1.00e+2 & +1.29e+1 & -7.26e+1 & -5.09e+1 & -4.68e+1 & -4.33e+1 & +2.11e+2 & +1.86e+2 & -3.03e+2 & -3.02e+2 & +1.73e+3 & -9.42e-1           \\ 
\midrule[0.3pt]
Term {\it w.r.t.} death & $\nu_{1,1}$ & $\nu_{2,1}$ & $\nu_{3,1}$ & $\nu_{4,1}$ & $\nu_{5,1}$ & $\nu_{6,1}$ & $\nu_{1,2}$ & $\nu_{2,2}$ & $\nu_{3,2}$ & $\nu_{4,2}$ & $\nu_{5,2}$ & $\nu_{6,2}$ & $\zeta_{1}$ & $\zeta_{2}$ \\ \hline
Coefficient & -1.92e+3 & -1.09e+3 & {\bf +3.61e+2} & {\bf +1.17e+3} & -3.12e+3 & {\bf +4.77e+3} & -1.55e+3 & -9.95e+2 & {\bf +2.54e+2} & {\bf +1.15e+3} & -3.19e+3 & {\bf +4.64e+3} & -1.68e+4 & -1.17e+3 \\
$p$-value   & +2.81e-6 & +4.01e-6 & +1.97e-5 & +7.78e-5 & +4.50e-5 & +9.21e-5 & +1.03e-5 & +3.21e-5 & +8.09e-7 & +3.14e-6 & +9.03e-7 & +9.67e-7 & +1.56e-7 & +6.85e-5 \\
$t$-value   & -1.11e+2 & -9.00e+1 & -6.31e+1 & -5.76e+1 & +2.09e+1 & +1.47e+1 & +6.80e+1 & +6.66e+1 & -1.80e+2 & -1.84e+2 & +2.76e+2 & +2.68e+2 & -9.71e+2 & -6.70e+1 \\
\midrule[0.3pt]\bottomrule[1pt]
\end{tabular}
}
\end{table}

\paragraph{Dependence on local covariates}

Table~\ref{tab:covariate-coeff} summarizes the fitted coefficients of local covariates in the model. 
The first and second rows indicate the corresponding time lag and the category of coefficients, respectively. The first 12 columns correspond to the community mobility, and the last two columns correspond to the demographic factors. Positive coefficients have been put in bold to highlight the positive correlation with cases or deaths. The coefficients can be compared across factors as the covariates are standardized first.
As we can see, most of the covariates are statistically significant, with small $p$-values ($< .05$) except for the proportion of the elderly population age 65 and older. The positive coefficients are in bold, which indicates a positive correlation between the covariates and the cases or deaths. In particular, we observe that, for the cases, the coefficients of mobility in \emph{workplaces}, \emph{retail, and recreation}, \emph{transit stations} have large positive values ($> 9 \times 10^2$), which indicates that the increase of mobility in these areas led to the rapid spread of the COVID-19. However, things are the opposite for the deaths, where the coefficients of mobility in \emph{grocery and  pharmacies}, \emph{parks}, \emph{residential} have large positive values ($> 2 \times 10^2$). Moreover, the population's coefficient for the cases is significantly larger than the other covariates, and it confirms that population density is the dominant factor in spreading the disease. Last, we have found the proportion of the elderly population is significantly related to the deaths and has no clear connection to the cases. 

\section{Discussion}
\label{sec:discussion}

While still in the development stages, the proposed spatio-temporal model has shown immense promise in modeling and predicting the deaths and confirmed cases of COVID-19 in the United States. Nevertheless, there remain numerous open questions and rooms for improvements. For example, the uncertainty in the count data commonly exists and can affect accuracy. It would be interesting to incorporate the serology data as an additional data source to calibrate our model. To avoid negative output, we may adapt the current problem into a Poisson regression with the log-linear model or apply a Rectified Linear Unit (ReLU) to the output to disallow the negative values. It assumes the response variable $\boldsymbol x_t$ has a Poisson distribution and assumes the logarithm of its expected value can be modeled by the linear model defined in \eqref{eq:case-death-model}. In particular, this adaption plays a vital role in predicting states with fewer confirmed cases and deaths, such as Hawaii and Delaware. 

\bibliographystyle{plain}
\bibliography{refs}

\begin{thebibliography}{10}

\bibitem{Adda2016}
J{\'e}r{\^o}me Adda.
\newblock Economic activity and the spread of viral diseases: Evidence from
  high frequency data.
\newblock {\em The Quarterly Journal of Economics}, 131(2):891--941, February
  2016.

\bibitem{agosto2020poisson}
Arianna Agosto and Paolo Giudici.
\newblock A poisson autoregressive model to understand covid-19 contagion
  dynamics.
\newblock {\em Risks}, 8(3):77, 2020.

\bibitem{alazab2020covid}
Moutaz Alazab, Albara Awajan, Abdelwadood Mesleh, Ajith Abraham, Vansh Jatana,
  and Salah Alhyari.
\newblock Covid-19 prediction and detection using deep learning.
\newblock {\em International Journal of Computer Information Systems and
  Industrial Management Applications}, 12:168--181, 2020.

\bibitem{altieri2020curating}
Nick Altieri, Rebecca~L Barter, James Duncan, Raaz Dwivedi, Karl Kumbier, Xiao
  Li, Robert Netzorg, Briton Park, Chandan Singh, Yan~Shuo Tan, Tiffany Tang,
  Yu~Wang, Chao Zhang, and Bin Yu.
\newblock Curating a {COVID}-19 data repository and forecasting county-level
  death counts in the united states.
\newblock {\em arXiv preprint arXiv:2005.07882}, 2020.

\bibitem{alzahrani2020forecasting}
Saleh~I Alzahrani, Ibrahim~A Aljamaan, and Ebrahim~A Al-Fakih.
\newblock Forecasting the spread of the covid-19 pandemic in saudi arabia using
  arima prediction model under current public health interventions.
\newblock {\em Journal of infection and public health}, 13(7):914--919, 2020.

\bibitem{Balcan2009}
Duygu Balcan, Vittoria Colizza, Bruno Gon{\c{c}}alves, Hao Hu, Jos{\'e}~J
  Ramasco, and Alessandro Vespignani.
\newblock Multiscale mobility networks and the spatial spreading of infectious
  diseases.
\newblock {\em Proceedings of the National Academy of Sciences},
  106(51):21484--21489, December 2009.

\bibitem{bertozzi2020challenges}
Andrea~L Bertozzi, Elisa Franco, George Mohler, Martin~B Short, and Daniel
  Sledge.
\newblock The challenges of modeling and forecasting the spread of covid-19.
\newblock {\em Proceedings of the National Academy of Sciences},
  117(29):16732--16738, 2020.

\bibitem{ACS2019}
The U.S.~Census Bureau.
\newblock American community survey, 2019.

\bibitem{caccavo2020chinese}
Diego Caccavo.
\newblock Chinese and italian covid-19 outbreaks can be correctly described by
  a modified sird model.
\newblock {\em medRxiv}, April 2020.

\bibitem{CDC}
{Centers for Disease Control and Prevention}.
\newblock {COVID}-19 forecasts: Deaths, 2021.

\bibitem{chakraborty2020theta}
Tanujit Chakraborty, Arinjita Bhattacharyya, and Monalisha Pattnaik.
\newblock Theta autoregressive neural network model for covid-19 outbreak
  predictions.
\newblock {\em medRxiv}, 2020.

\bibitem{chiang2020hawkes}
Wen-Hao Chiang, Xueying Liu, and George Mohler.
\newblock Hawkes process modeling of covid-19 with mobility leading indicators
  and spatial covariates.
\newblock {\em medRxiv}, 2020.

\bibitem{Chinazzi2020}
Matteo Chinazzi, Jessica~T Davis, Marco Ajelli, Corrado Gioannini, Maria
  Litvinova, Stefano Merler, Ana~Pastore y~Piontti, Kunpeng Mu, Luca Rossi,
  Kaiyuan Sun, et~al.
\newblock The effect of travel restrictions on the spread of the 2019 novel
  coronavirus (covid-19) outbreak.
\newblock {\em Science}, 368(6489):395--400, April 2020.

\bibitem{Colizza2006}
Vittoria Colizza, Alain Barrat, Marc Barth{\'e}lemy, and Alessandro Vespignani.
\newblock The role of the airline transportation network in the prediction and
  predictability of global epidemics.
\newblock {\em Proceedings of the National Academy of Sciences},
  103(7):2015--2020, February 2006.

\bibitem{CCDCP2020}
Chinese Center for Disease~Control Epidemiology Working Group~for NCIP
  Epidemic~Response and Prevention.
\newblock The epidemiological characteristics of an outbreak of 2019 novel
  coronavirus diseases (covid-19) in china.
\newblock {\em Zhonghua liu xing bing xue za zhi = Zhonghua liuxingbingxue
  zazhi}, 41(2):145—151, February 2020.

\bibitem{fadly2020approach}
Ferdian Fadly and Erika Sari.
\newblock An approach to measure the death impact of covid-19 in jakarta using
  autoregressive integrated moving average (arima).
\newblock {\em Unnes Journal of Public Health}, 9(2):108--116, 2020.

\bibitem{Fang2009}
Li-Qun Fang, Sake~J De~Vlas, Dan Feng, Song Liang, You-Fu Xu, Jie-Ping Zhou,
  Jan~Hendrik Richardus, and Wu-Chun Cao.
\newblock Geographical spread of sars in mainland china.
\newblock {\em Tropical Medicine \& International Health}, 14(s1):14--20,
  October 2009.

\bibitem{fernandez2020estimating}
Jes{\'u}s Fern{\'a}ndez-Villaverde and Charles~I Jones.
\newblock Estimating and simulating a sird model of covid-19 for many
  countries, states, and cities.
\newblock Working Paper 27128, National Bureau of Economic Research, May 2020.

\bibitem{Gaetan2010}
Carlo Gaetan and Xavier Guyon.
\newblock {\em Spatial statistics and modeling}, volume~90.
\newblock Springer, New York, NY, USA, 2010.

\bibitem{ghosh2020covid}
Palash Ghosh, Rik Ghosh, and Bibhas Chakraborty.
\newblock Covid-19 in india: Statewise analysis and prediction.
\newblock {\em JMIR public health and surveillance}, 6(3):e20341, 2020.

\bibitem{Covidcommunity2020}
Google.
\newblock Covid-19 community mobility reports, 2020.

\bibitem{Harko2014}
Tiberiu Harko, Francisco~SN Lobo, and MK~Mak.
\newblock Exact analytical solutions of the susceptible-infected-recovered
  (sir) epidemic model and of the sir model with equal death and birth rates.
\newblock {\em Applied Mathematics and Computation}, 236:184--194, June 2014.

\bibitem{havers2020seroprevalence}
Fiona~P Havers, Carrie Reed, Travis Lim, Joel~M Montgomery, John~D Klena,
  Aron~J Hall, Alicia~M Fry, Deborah~L Cannon, Cheng-Feng Chiang, Aridth
  Gibbons, et~al.
\newblock Seroprevalence of antibodies to sars-cov-2 in 10 sites in the united
  states, march 23-may 12, 2020.
\newblock {\em JAMA Internal Medicine}, July 2020.

\bibitem{hawas2020generated}
Mohamed Hawas.
\newblock Generated time-series prediction data of covid-19' s daily infections
  in brazil by using recurrent neural networks.
\newblock {\em Data in brief}, 32:106175, 2020.

\bibitem{Hethcote2000}
Herbert~W Hethcote.
\newblock The mathematics of infectious diseases.
\newblock {\em SIAM review}, 42(4):599--653, October 2000.

\bibitem{hou2020effectiveness}
Can Hou, Jiaxin Chen, Yaqing Zhou, Lei Hua, Jinxia Yuan, Shu He, Yi~Guo, Sheng
  Zhang, Qiaowei Jia, Chenhui Zhao, Jing Zhang, Guangxu Xu, and Enzhi Jia.
\newblock The effectiveness of quarantine of wuhan city against the corona
  virus disease 2019 (covid-19): A well-mixed seir model analysis.
\newblock {\em Journal of medical virology}, 92(7):841--848, April 2020.

\bibitem{IHME}
{Institute for Health Metrics and Evaluation}.
\newblock {COVID}-19 projections for the united states, 2021.

\bibitem{covid2021modeling}
{Institute for Health Metrics and Evaluation}.
\newblock Modeling covid-19 scenarios for the united states.
\newblock {\em Nature medicine}, 27(1):94, 2021.

\bibitem{Jia2020}
Jayson~S Jia, Xin Lu, Yun Yuan, Ge~Xu, Jianmin Jia, and Nicholas~A Christakis.
\newblock Population flow drives spatio-temporal distribution of covid-19 in
  china.
\newblock {\em Nature}, 582(7812):389--394, April 2020.

\bibitem{Kang2020}
Dayun Kang, Hyunho Choi, Jong-Hun Kim, and Jungsoon Choi.
\newblock Spatial epidemic dynamics of the covid-19 outbreak in china.
\newblock {\em International Journal of Infectious Diseases}, 94:96--102, May
  2020.

\bibitem{kapoor2020examining}
Amol Kapoor, Xue Ben, Luyang Liu, Bryan Perozzi, Matt Barnes, Martin Blais, and
  Shawn O'Banion.
\newblock Examining covid-19 forecasting using spatio-temporal graph neural
  networks.
\newblock {\em arXiv preprint arXiv:2007.03113}, 2020.

\bibitem{khan2020modelling}
Firdos Khan, Alia Saeed, and Shaukat Ali.
\newblock Modelling and forecasting of new cases, deaths and recover cases of
  covid-19 by using vector autoregressive model in pakistan.
\newblock {\em Chaos, Solitons \& Fractals}, 140:110189, 2020.

\bibitem{killian2020evaluating}
Jackson~A Killian, Marie Charpignon, Bryan Wilder, Andrew Perrault, Milind
  Tambe, and Maimuna~S Majumder.
\newblock Evaluating covid-19 lockdown and business-sector-specific reopening
  policies for three us states, May 2020.

\bibitem{kirbacs2020comparative}
{\.I}smail K{\i}rba{\c{s}}, Adnan S{\"o}zen, Azim~Do{\u{g}}u{\c{s}} Tuncer, and
  Fikret~{\c{S}}inasi Kazanc{\i}o{\u{g}}lu.
\newblock Comparative analysis and forecasting of covid-19 cases in various
  european countries with arima, narnn and lstm approaches.
\newblock {\em Chaos, Solitons \& Fractals}, 138:110015, 2020.

\bibitem{kou2020unmasking}
SC~Kou, Shihao Yang, Chia-Jung Chang, Teck-Hua Ho, and Lisa Graver.
\newblock Unmasking the actual covid-19 case count.
\newblock {\em Clinical Infectious Diseases}, May 2020.

\bibitem{lai2020effect}
Shengjie Lai, Nick~W Ruktanonchai, Liangcai Zhou, Olivia Prosper, Wei Luo,
  Jessica~R Floyd, Amy Wesolowski, Mauricio Santillana, Chi Zhang, Xiangjun Du,
  et~al.
\newblock Effect of non-pharmaceutical interventions to contain covid-19 in
  china.
\newblock {\em Nature}, May 2020.

\bibitem{Lee2014}
Mi~Lim Lee, David Goldsman, Seong-Hee Kim, and Kwok-Leung Tsui.
\newblock Spatiotemporal biosurveillance with spatial clusters: control limit
  approximation and impact of spatial correlation.
\newblock {\em IIE Transactions}, 46(8):813--827, May 2014.

\bibitem{Liu2020}
Lu~Liu.
\newblock Emerging study on the transmission of the novel coronavirus
  (covid-19) from urban perspective: Evidence from china.
\newblock {\em Cities}, 103:102759, August 2020.

\bibitem{LANL}
{Los Alamos National Laboratory}.
\newblock {LANL} {COVID}-19 cases and deaths forecasts, 2021.

\bibitem{lu2020estimating}
Fred~S Lu, Andre~T Nguyen, Nick Link, and Mauricio Santillana.
\newblock Estimating the prevalence of covid-19 in the united states: Three
  complementary approaches.
\newblock {\em medRxiv}, 2020.

\bibitem{maleki2020time}
Mohsen Maleki, Mohammad~Reza Mahmoudi, Darren Wraith, and Kim-Hung Pho.
\newblock Time series modelling to forecast the confirmed and recovered cases
  of covid-19.
\newblock {\em Travel medicine and infectious disease}, 37:101742, 2020.

\bibitem{Meng2005}
Bin Meng, Jinfeng Wang, J~Liu, J~Wu, and E~Zhong.
\newblock Understanding the spatial diffusion process of severe acute
  respiratory syndrome in beijing.
\newblock {\em Public Health}, 119(12):1080--1087, December 2005.

\bibitem{deathrate}
U.S. News.
\newblock {COVID-19} deaths per 100k, 2021.

\bibitem{MOBS}
{Northeastern University, Laboratory for the Modeling of Biological and
  Socio-technical Systems}.
\newblock {COVID}-19 modeling, 2021.

\bibitem{onder2020case}
Graziano Onder, Giovanni Rezza, and Silvio Brusaferro.
\newblock Case-fatality rate and characteristics of patients dying in relation
  to covid-19 in italy.
\newblock {\em Jama}, 323(18):1775--1776, March 2020.

\bibitem{poirier2020real}
Canelle Poirier, Dianbo Liu, Leonardo Clemente, Xiyu Ding, Matteo Chinazzi,
  Jessica Davis, Alessandro Vespignani, and Mauricio Santillana.
\newblock Real-time forecasting of the covid-19 outbreak in chinese provinces:
  Machine learning approach using novel digital data and estimates from
  mechanistic models.
\newblock {\em Journal of medical Internet research}, 22(8):e20285, 2020.

\bibitem{poirier2020role}
Canelle Poirier, Wei Luo, Maimuna~S Majumder, Dianbo Liu, Kenneth Mandl, Todd
  Mooring, and Mauricio Santillana.
\newblock The role of environmental factors on transmission rates of the
  covid-19 outbreak: An initial assessment in two spatial scales.
\newblock {\em Available at SSRN: https://ssrn.com/abstract=3552677}, March
  2020.

\bibitem{Covidproj2020}
The COVID~Tracking Project.
\newblock About the project, 2020.

\bibitem{ray2020ensemble}
Evan~L Ray, Nutcha Wattanachit, Jarad Niemi, Abdul~Hannan Kanji, Katie House,
  Estee~Y Cramer, Johannes Bracher, Andrew Zheng, Teresa~K Yamana, Xinyue
  Xiong, et~al.
\newblock Ensemble forecasts of coronavirus disease 2019 (covid-19) in the
  {US}.
\newblock {\em MedRXiv}, 2020.

\bibitem{roy2020spatial}
Santanu Roy, Gouri~Sankar Bhunia, and Pravat~Kumar Shit.
\newblock Spatial prediction of covid-19 epidemic using arima techniques in
  india.
\newblock {\em Modeling earth systems and environment}, pages 1--7, 2020.

\bibitem{saba2020forecasting}
Amal~I Saba and Ammar~H Elsheikh.
\newblock Forecasting the prevalence of covid-19 outbreak in egypt using
  nonlinear autoregressive artificial neural networks.
\newblock {\em Process safety and environmental protection}, 141:1--8, 2020.

\bibitem{Sajadi2020}
Mohammad~M Sajadi, Parham Habibzadeh, Augustin Vintzileos, Shervin Shokouhi,
  Fernando Miralles-Wilhelm, and Anthony Amoroso.
\newblock Temperature and latitude analysis to predict potential spread and
  seasonality for covid-19.
\newblock {\em Available at SSRN: https://ssrn.com/abstract=3550308}, March
  2020.

\bibitem{singh2020prediction}
Ram~Kumar Singh, Meenu Rani, Akshaya~Srikanth Bhagavathula, Ranjit Sah,
  Alfonso~J Rodriguez-Morales, Himangshu Kalita, Chintan Nanda, Shashi Sharma,
  Yagya~Datt Sharma, Ali~A Rabaan, et~al.
\newblock Prediction of the covid-19 pandemic for the top 15 affected
  countries: Advanced autoregressive integrated moving average (arima) model.
\newblock {\em JMIR public health and surveillance}, 6(2):e19115, 2020.

\bibitem{singhal2020review}
Tanu Singhal.
\newblock A review of coronavirus disease-2019 (covid-19).
\newblock {\em The Indian Journal of Pediatrics}, 87(4):281--286, March 2020.

\bibitem{tamang2020forecasting}
SK~Tamang, PD~Singh, and B~Datta.
\newblock Forecasting of covid-19 cases based on prediction using artificial
  neural network curve fitting technique.
\newblock {\em Global Journal of Environmental Science and Management},
  6(Special Issue (Covid-19)):53--64, 2020.

\bibitem{UTaustin}
{The University of Texas COVID-19 Modeling Consortium}.
\newblock {COVID}-19 mortality projections for {US} states, 2021.

\bibitem{NYT2019}
The New~York Times.
\newblock We’re sharing coronavirus case data for every u.s. county, 2020.

\bibitem{woody2020projections}
Spencer Woody, Mauricio~Garcia Tec, Maytal Dahan, Kelly Gaither, Michael
  Lachmann, Spencer Fox, Lauren~Ancel Meyers, and James~G Scott.
\newblock Projections for first-wave covid-19 deaths across the us using
  social-distancing measures derived from mobile phones.
\newblock {\em medRxiv}, 2020.

\bibitem{yang2020modified}
Zifeng Yang, Zhiqi Zeng, Ke~Wang, Sook-San Wong, Wenhua Liang, Mark Zanin, Peng
  Liu, Xudong Cao, Zhongqiang Gao, Zhitong Mai, et~al.
\newblock Modified seir and ai prediction of the epidemics trend of covid-19 in
  china under public health interventions.
\newblock {\em Journal of Thoracic Disease}, 12(3):165--174, March 2020.

\bibitem{zhao2020well}
Zhuowen Zhao, Kieran Nehil-Puleo, and Yangzhi Zhao.
\newblock How well can we forecast the covid-19 pandemic with curve fitting and
  recurrent neural networks?
\newblock {\em medRxiv}, 2020.

\end{thebibliography}

\appendix

\section{Other numerical results}
\label{append:other-results}

\begin{figure}[!ht]
\centering
\begin{subfigure}[h]{.325\linewidth}
\includegraphics[width=\textwidth]{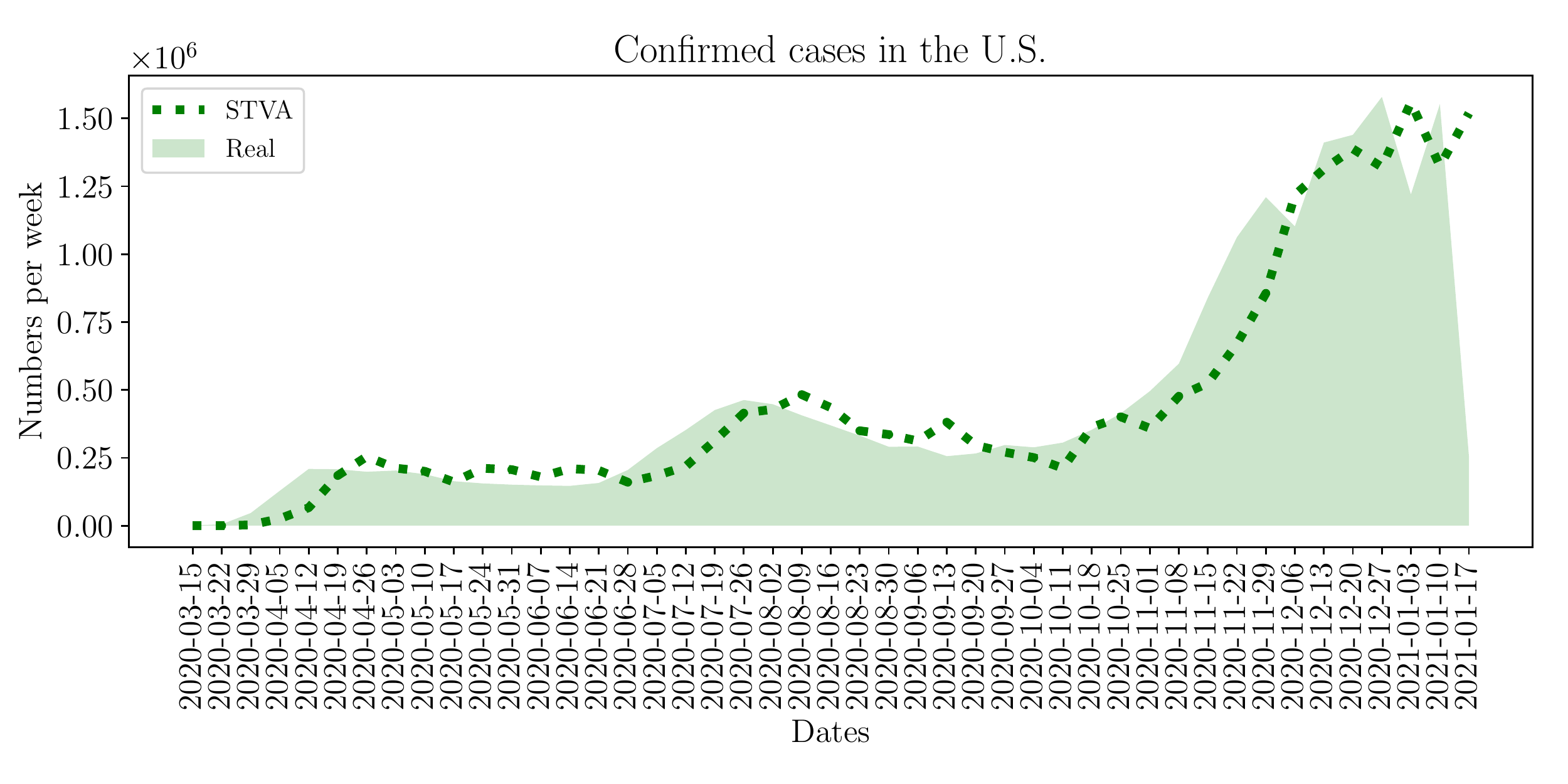}
\end{subfigure}
\begin{subfigure}[h]{.325\linewidth}
\includegraphics[width=\textwidth]{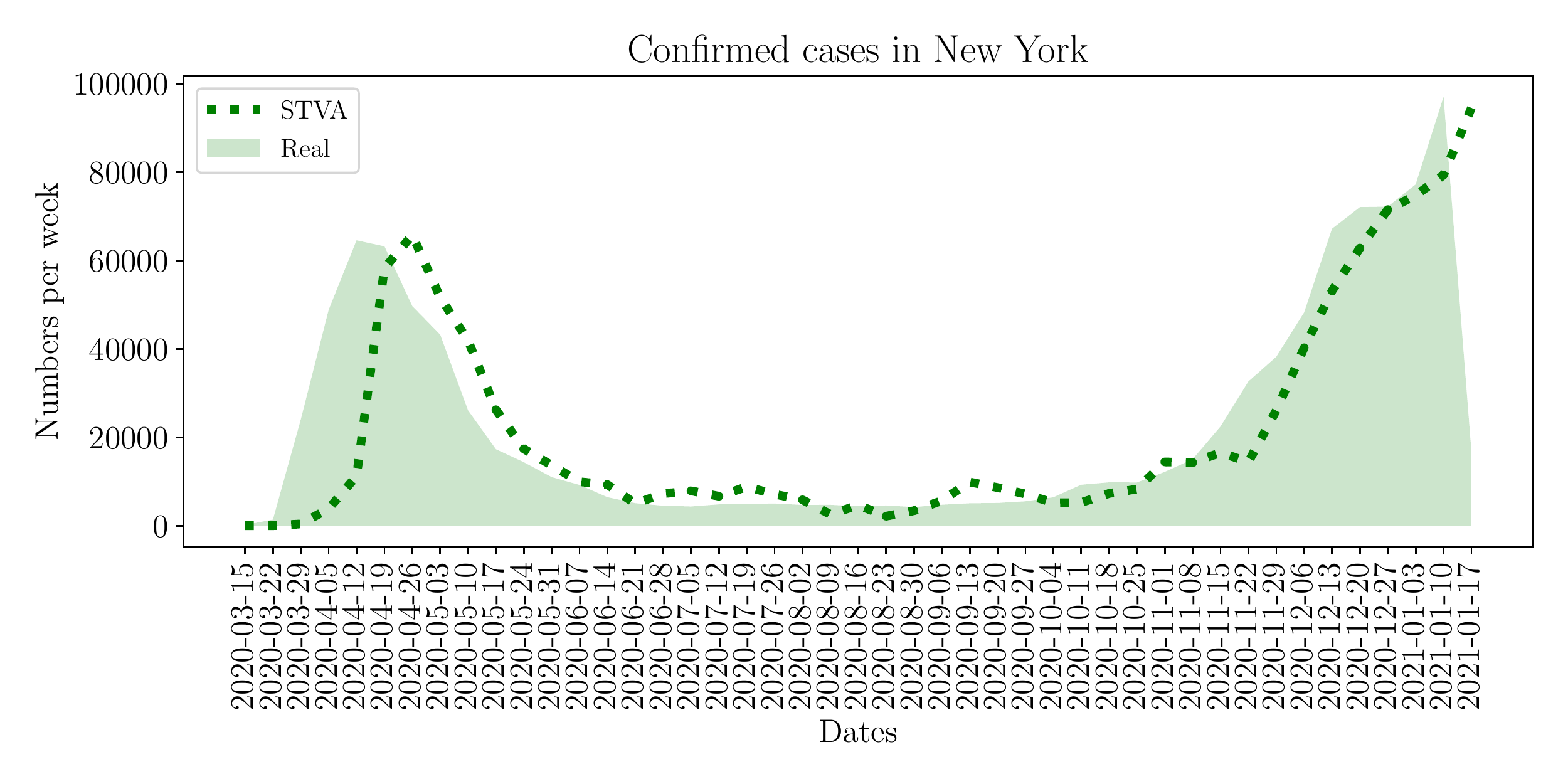}
\end{subfigure}
\begin{subfigure}[h]{.325\linewidth}
\includegraphics[width=\textwidth]{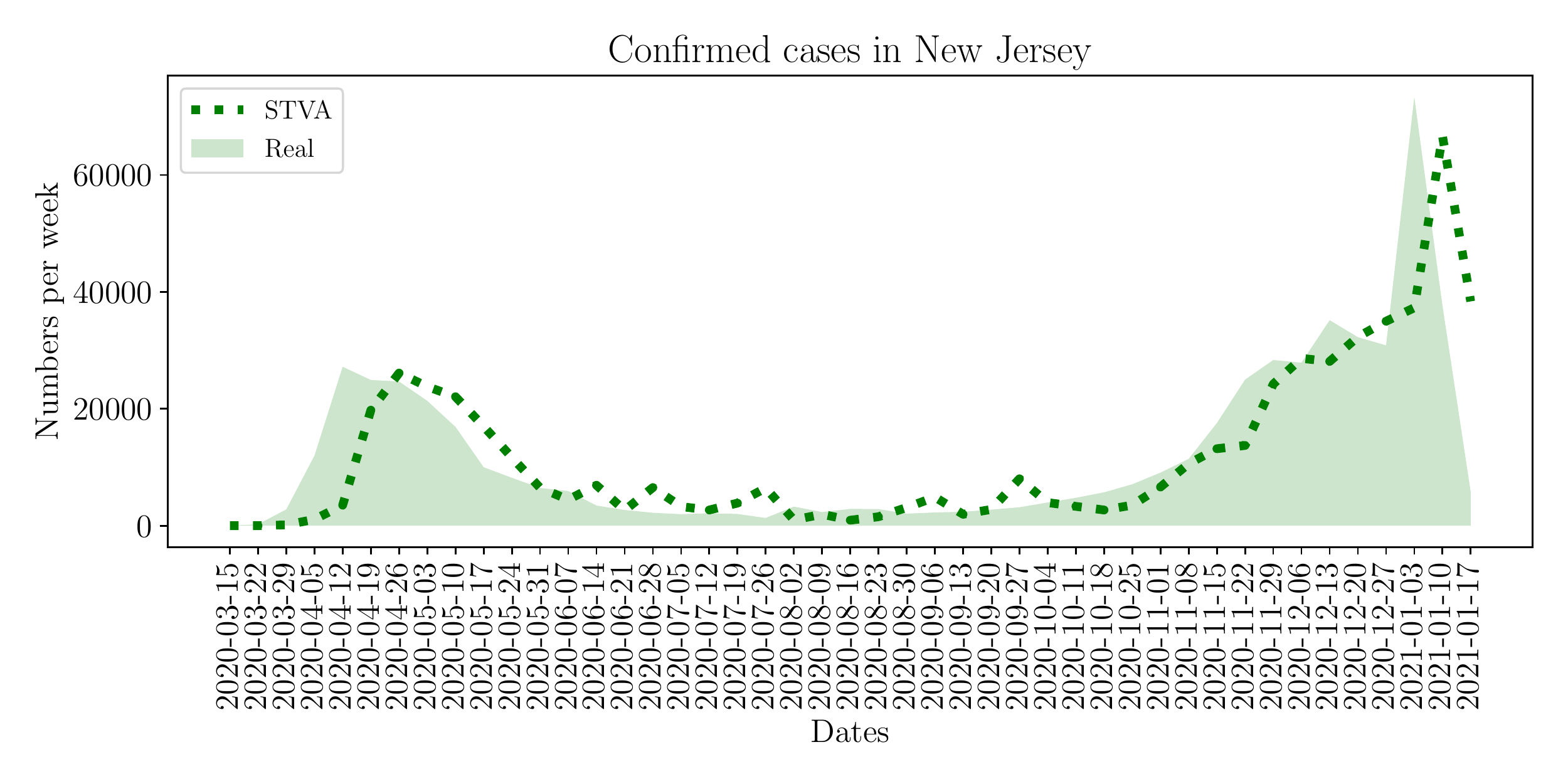}
\end{subfigure}
\vfill
\begin{subfigure}[h]{.325\linewidth}
\includegraphics[width=\textwidth]{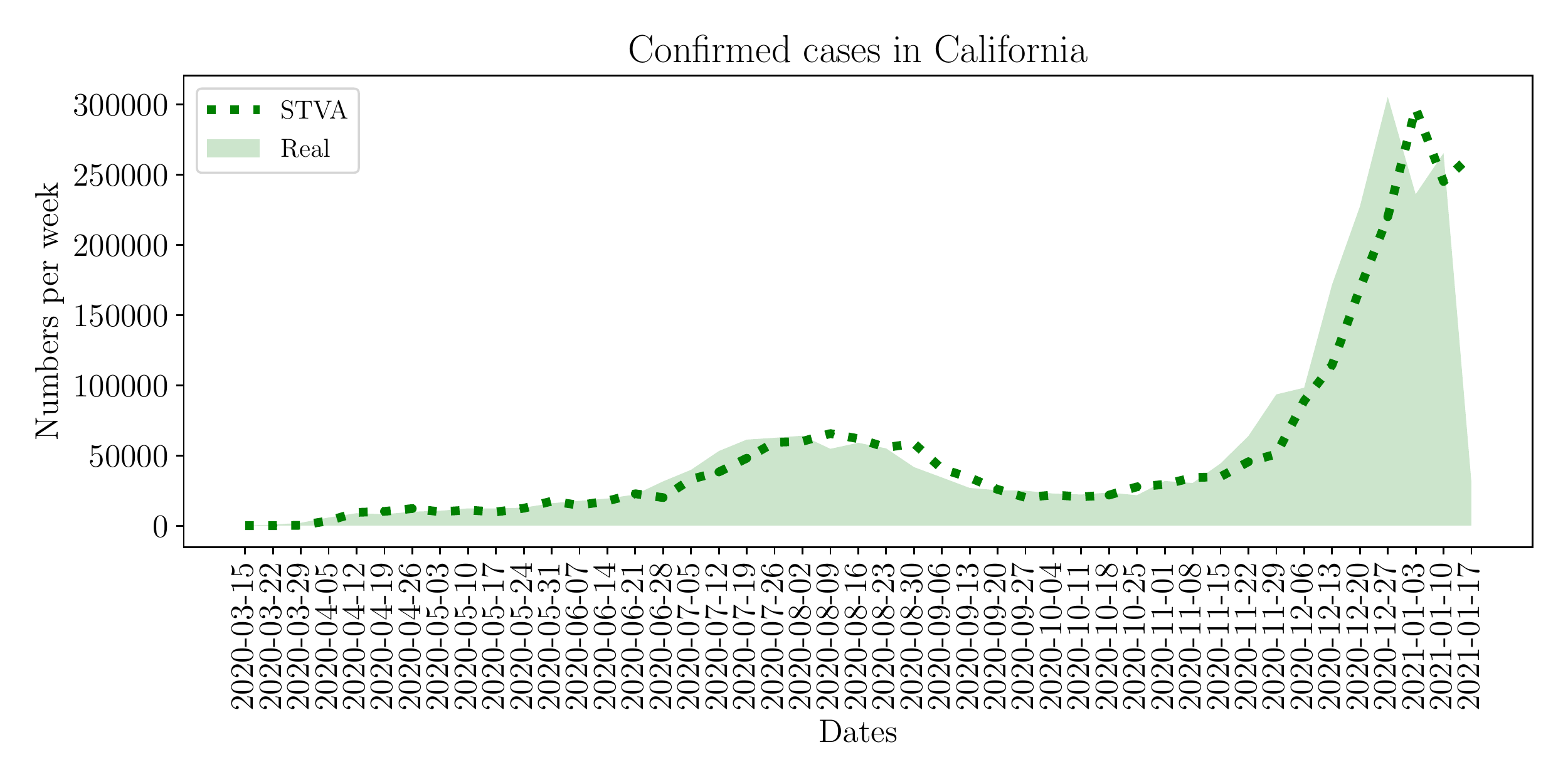}
\end{subfigure}
\begin{subfigure}[h]{.325\linewidth}
\includegraphics[width=\textwidth]{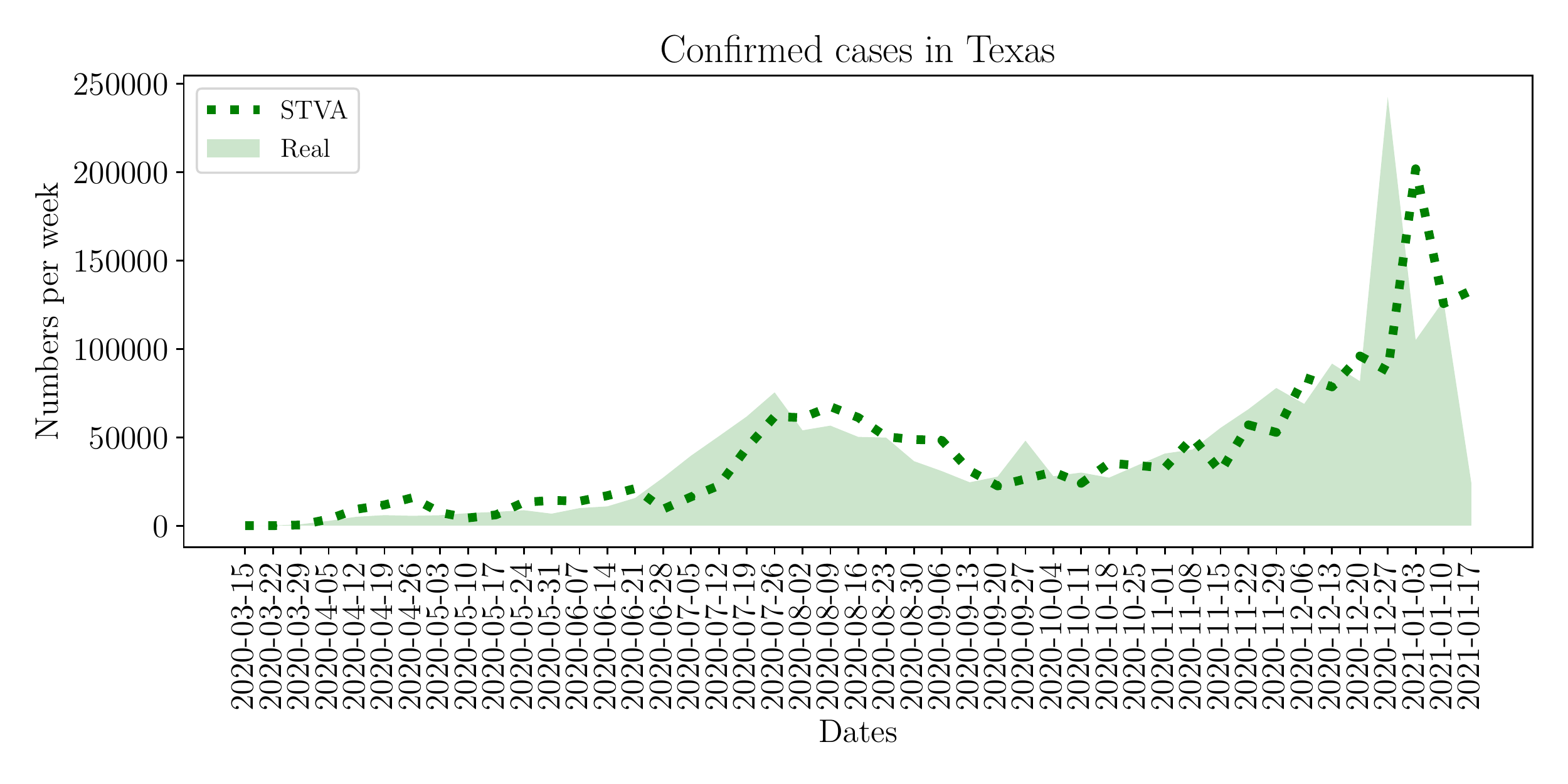}
\end{subfigure}
\begin{subfigure}[h]{.325\linewidth}
\includegraphics[width=\textwidth]{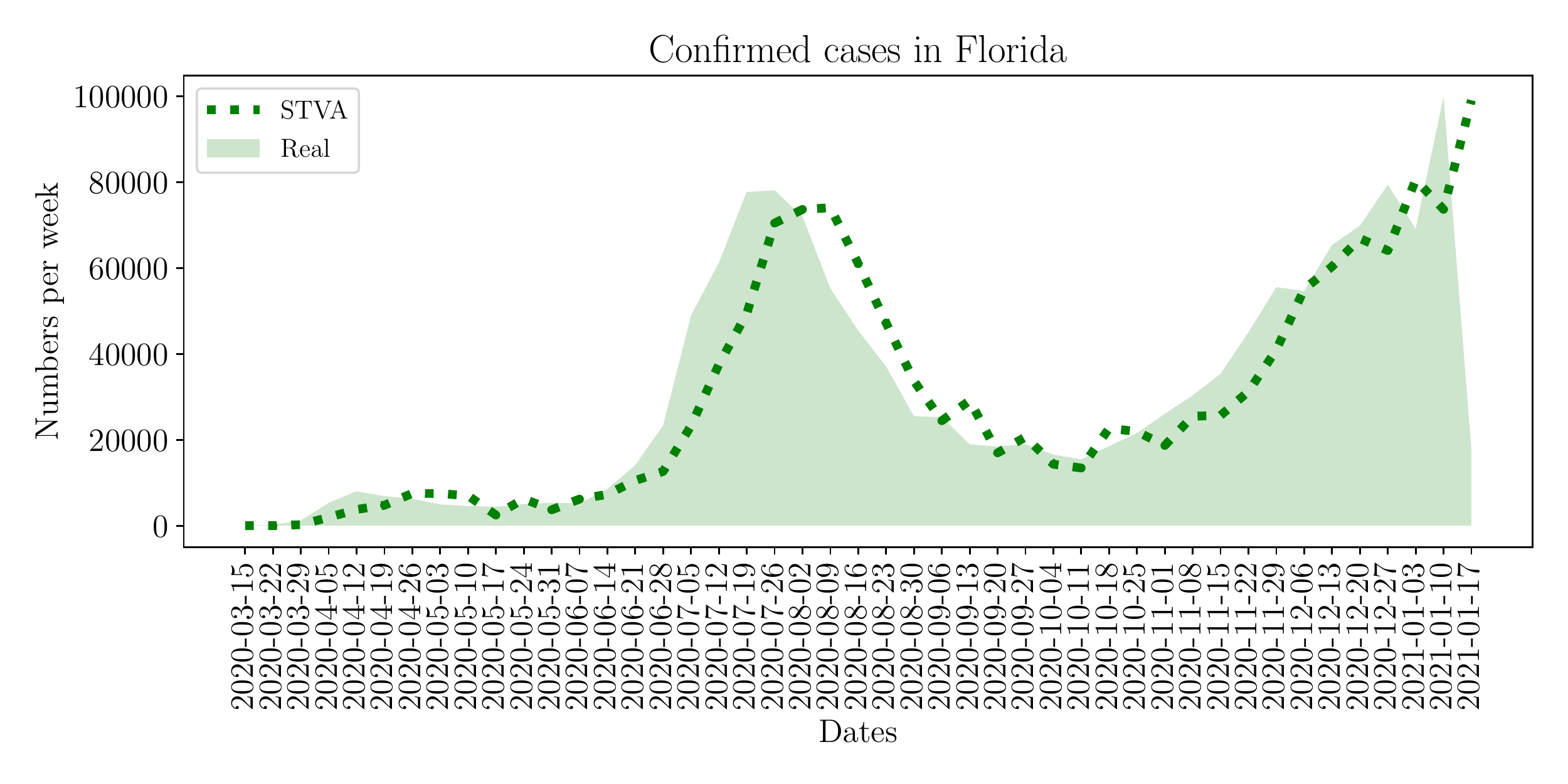}
\end{subfigure}
\vfill
\begin{subfigure}[h]{.325\linewidth}
\includegraphics[width=\textwidth]{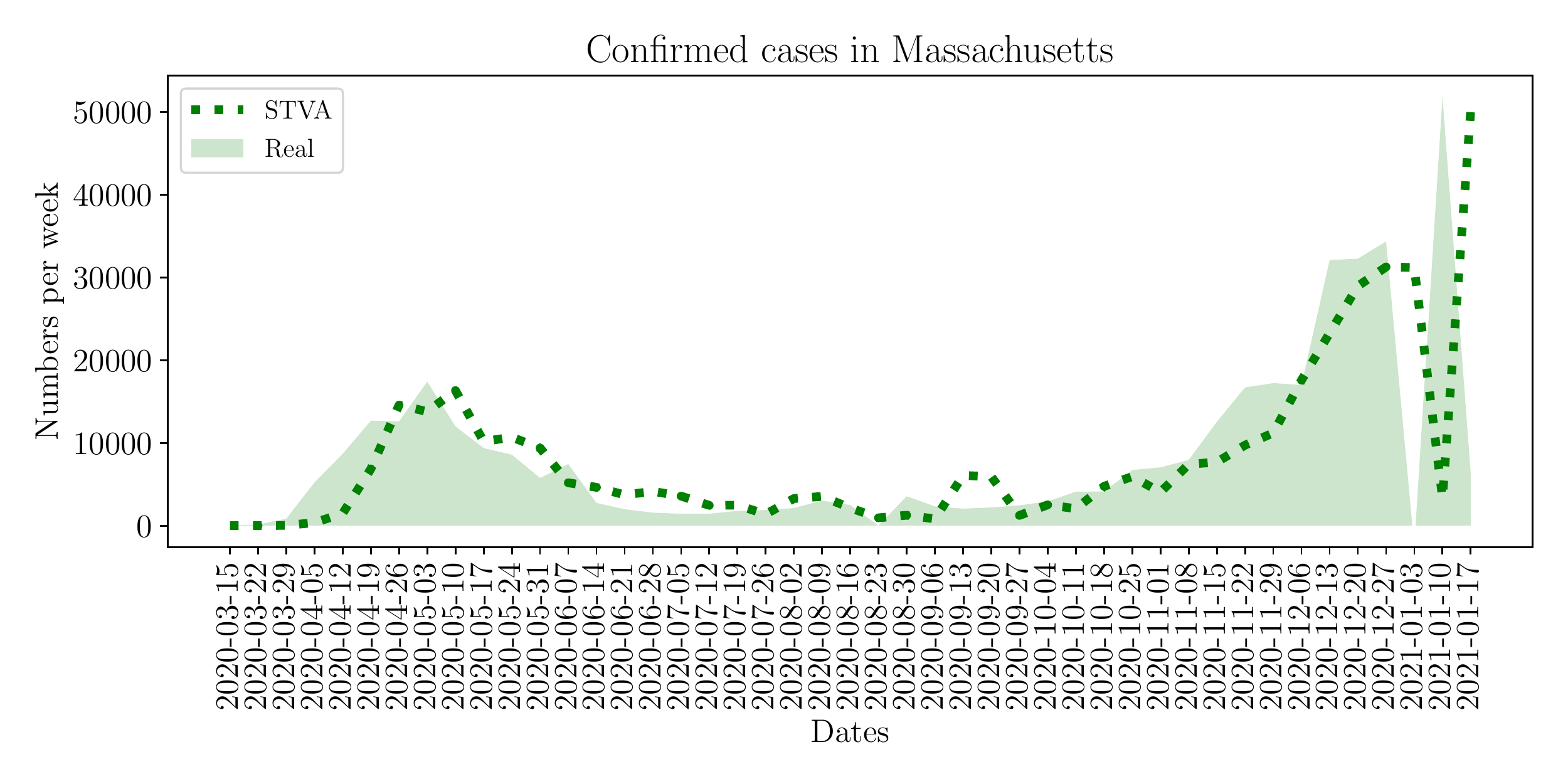}
\end{subfigure}
\begin{subfigure}[h]{.325\linewidth}
\includegraphics[width=\textwidth]{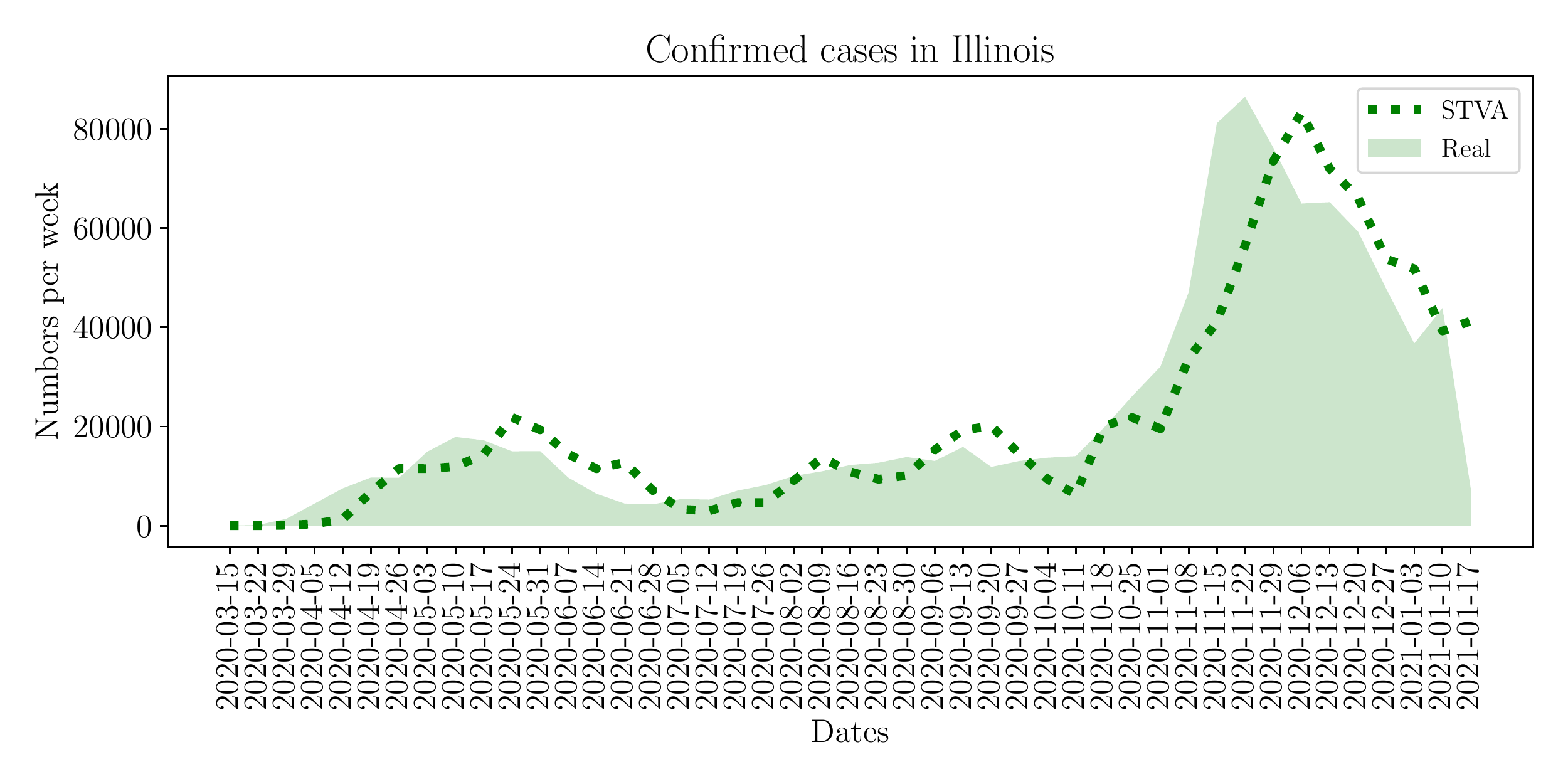}
\end{subfigure}
\begin{subfigure}[h]{.325\linewidth}
\includegraphics[width=\textwidth]{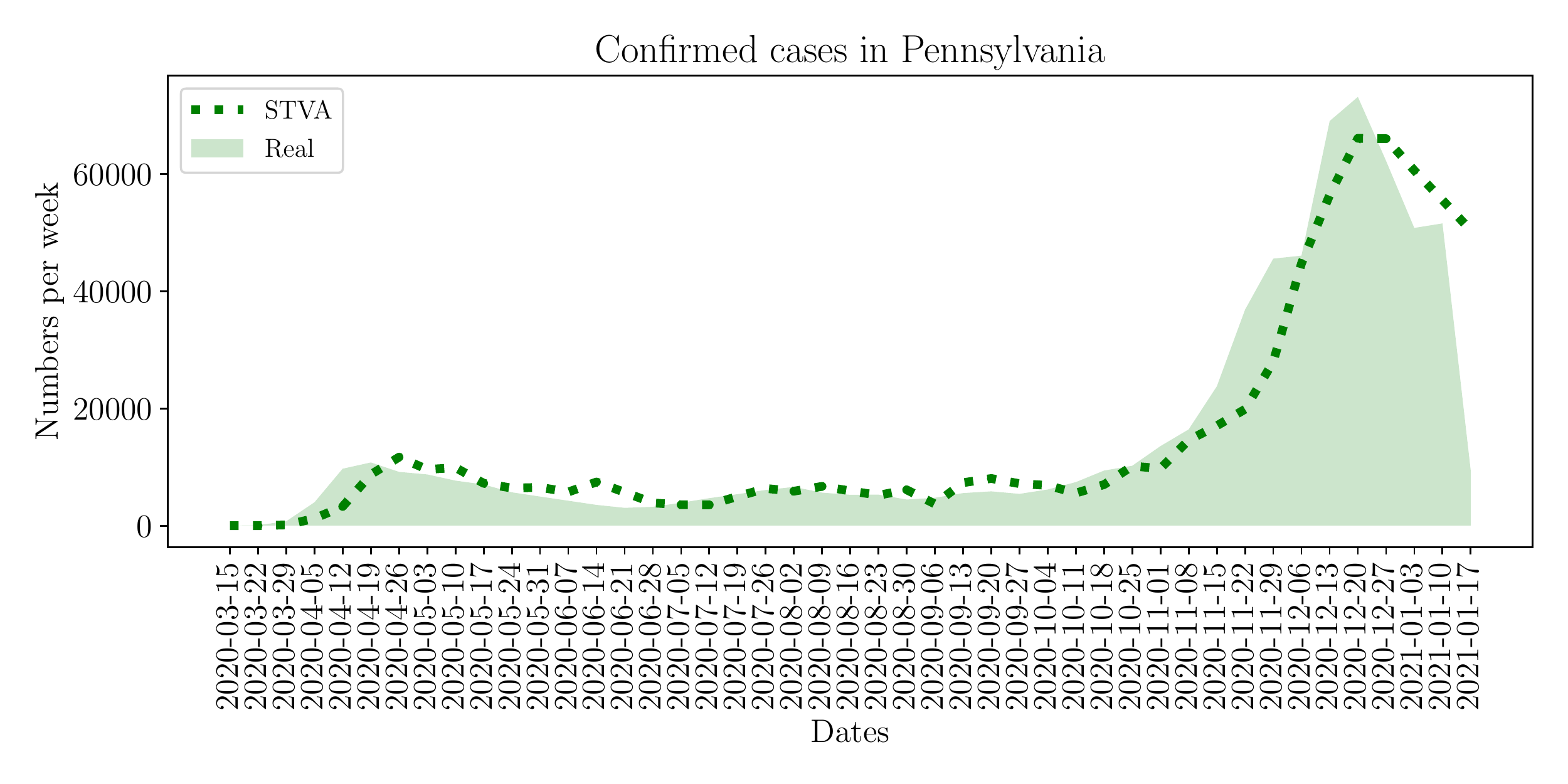}
\end{subfigure}
\caption{In-sample estimated cases (green dotted lines) for the U.S. and other eight major states with the highest number of COVID-19 deaths in the U.S.. Figures are sorted in descending order of the total number of deaths since March 15th, 2020.}
\label{fig:statewise-case}
\end{figure}

\end{document}